\documentclass[twocolumn]{aa}  
\usepackage{longtable}
\usepackage{pdflscape} 
\usepackage{lscape}
\usepackage{xcolor}
\usepackage{listings}
\usepackage{natbib}
\bibpunct{(}{)}{;}{a}{}{,} 
\usepackage{booktabs,caption}
\usepackage[flushleft]{threeparttable}
\usepackage{adjustbox}
\usepackage{dcolumn}
\newcommand\gaia{\textit{Gaia}}
\newcommand{\gdr}[1]{\gaia~DR#1}

\newcommand\kms{\ensuremath{\text{km~s}^{-1}}}

\newcommand\vlos{\ensuremath{v_\mathrm{los}}}

\newcommand\pmdec{\ensuremath{\mu_{\delta}}}
\newcommand\feh{\ensuremath{[\mathrm{Fe}/\mathrm{H}]}}

\definecolor{codegreen}{rgb}{0,0.6,0}
\definecolor{codegray}{rgb}{0.5,0.5,0.5}
\definecolor{codepurple}{rgb}{0.58,0,0.82}
\definecolor{backcolour}{rgb}{0.95,0.95,0.92}

\usepackage[normalem]{ulem}

\lstdefinestyle{mystyle}{
    commentstyle=\color{codegreen},
    keywordstyle=\color{magenta},
    numberstyle=\tiny\color{codegray},
    stringstyle=\color{codepurple},
    basicstyle=\ttfamily,
    breakatwhitespace=false,         
    breaklines=true,                 
    captionpos=b,                    
    keepspaces=true,                 
    numbers=none,                    
    showspaces=false,                
    showstringspaces=false,
    showtabs=false,                  
    tabsize=2
}
\title{A 3D view of dwarf galaxies with \gaia\ and VLT/FLAMES\\II. The Sextans dwarf spheroidal\thanks{Tables E.1 and E.2 are only available online via CDS}$^,$\thanks{Based on VLT/FLAMES observations collected at the European Organisation for Astronomical Research (ESO) in the Southern Hemisphere under programmes: 171.B-0588; 60.A-9800;  0102.B-0786}}

\titlerunning{A 3D view of Sextans}
\authorrunning{Tolstoy, Battaglia et al.}
\author{ Eline Tolstoy, \inst{1}   Giuseppina Battaglia, \inst{2, 3}  Jos\'{e} Mar\'{i}a Arroyo-Polonio, \inst{2, 3}  Anthony G.A. Brown\inst{4}, \\ Thom van Essen, \inst{1}, Davide Massari, \inst{5, 1} \'Asa Sk\'ulad\'ottir, \inst{6} Michael J. Irwin, \inst{7}  Salvatore Taibi, \inst{8} John Pritchard \inst{9} 
}
\institute{Kapteyn Astronomical Institute, University of Groningen, PO Box 800,
             9700AV Groningen, the Netherlands\\ \email{etolstoy@astro.rug.nl}
             \and
             Instituto de Astrof\'isica de Canarias, Calle V\'ia L\'actea s/n, E-38206 La Laguna, Tenerife, Spain
             \and 
             Universidad de La Laguna, Avda. Astrof\'isico Fco. S\'anchez, E-38205 La Laguna, Tenerife, Spain
             \and
             Leiden Observatory, Leiden University, Einsteinweg 55, NL-2333 CC Leiden, the Netherlands
             \and
             INAF - Osservatorio di Astrofisica e Scienza dello Spazio di Bologna, Via Gobetti 93/3, I-40129 Bologna, Italy
            \and
             Dipartimento di Fisica e Astronomia, Universit\'a degli Studi di Firenze, Via G. Sansone 1, I-50019 Sesto Fiorentino, Italy 
             \and
              Institute of Astronomy, Madingley Road, Cambridge CB3 0HA, UK
             \and
             Institute of Physics, Laboratory of Astrophysics, Ecole Polytechnique Fédérale de Lausanne (EPFL), 1290 Sauverny, Switzerland
            \and
              European Southern Observatory, Karl-Schwarzschild-Str. 2, D-85748 Garching bei M\"unchen, Germany
             }

\abstract{The Sextans dwarf spheroidal galaxy has been challenging to study in a comprehensive way as it is highly extended on the sky, with an uncertain but large tidal radius of between $80$--$160$ arcminutes (or $3$--$4$kpc), and an extremely low central surface brightness of $\Sigma_V \sim 26.2$ mag/arcsec$^2$. 
Here we present a new homogeneous survey of 41 VLT/FLAMES multi-fibre spectroscopic pointings that contain 2108 individual spectra, and combined with \gdr{3} photometry and astrometry we present \vlos\ measurements for 333    
individual Red Giant Branch stars that are consistent with membership in the Sextans dwarf spheroidal galaxy. In addition, we provide the metallicity, [Fe/H], determined from the two strongest Ca~II triplet lines, for 312 of these stars. 
We look again at the global characteristics of Sextans, deriving a mean line-of-sight velocity of $\langle$\vlos$\rangle = +227.1$km/s and a mean 
metallicity of 
$\langle$\feh$\rangle = -2.37$. The metallicity distribution is clearly double peaked, with the highest peak at [Fe/H]~$= -2.81$  and another broader peak at [Fe/H]~$= -2.09$. Thus it appears that Sextans hosts two populations and the superposition leads to a radial variation in the mean metallicity, with the  more metal rich population being centrally concentrated. 
In addition there is an intriguing group of 9 probable members in the outer region of Sextans at higher [Fe/H] than the mean in this region. If this group could be confirmed as members they would eliminate the metallicity gradient.
We also look again at the Colour-Magnitude Diagram
of the resolved stellar population in Sextans. We also look again at the relation between Sextans and the intriguingly nearby globular cluster, Pal~3.
The global properties of Sextans have not changed significantly compared to previous studies, but they are now more precise, and the sample of known members in the outer regions is now more complete.
}


\begin{document}
\flushbottom
\maketitle
\thispagestyle{empty}

\begin{figure*}[t]
\centering
\includegraphics[width=0.9\linewidth]{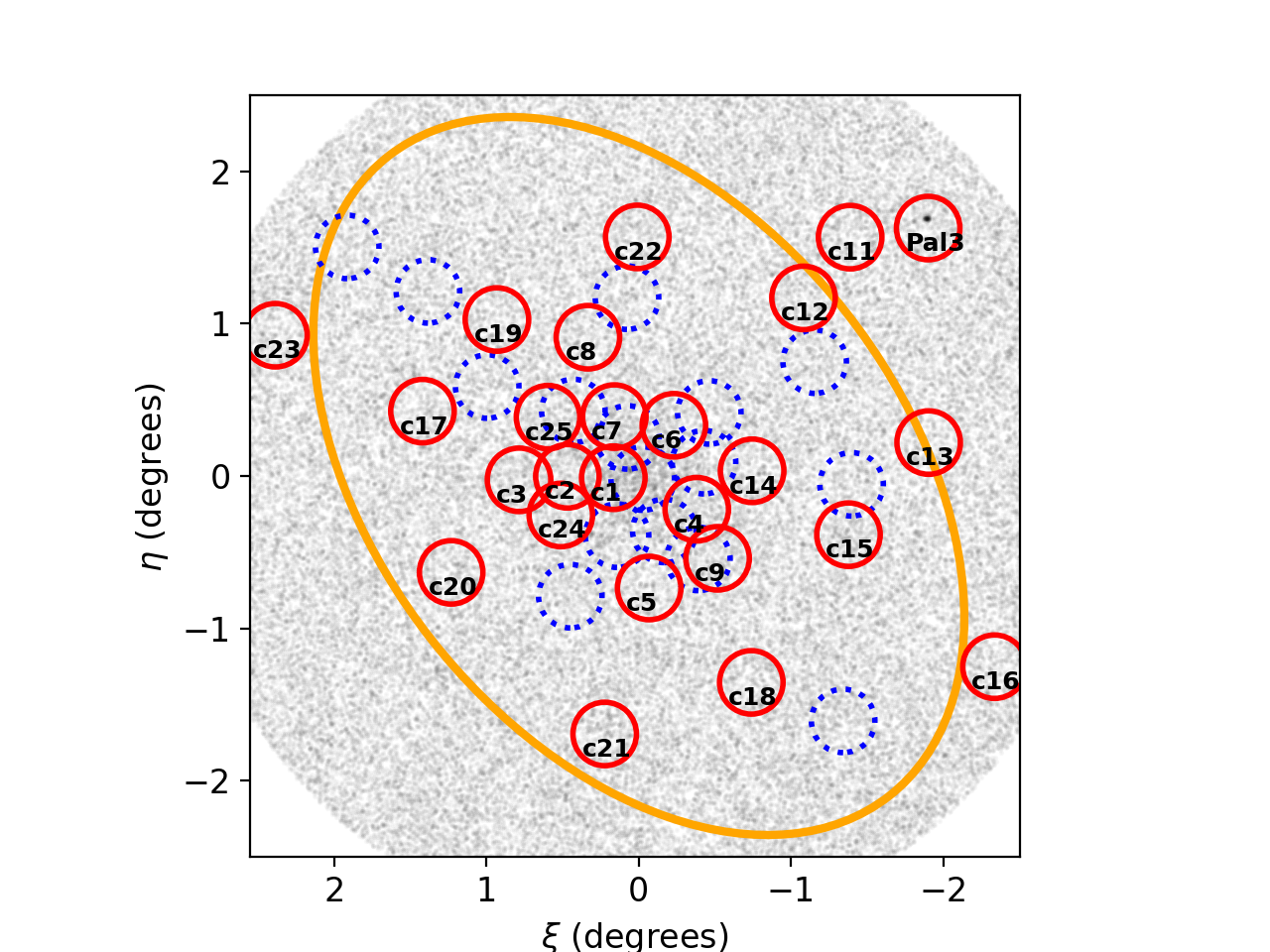}
\caption{The \gdr{3} catalogue  within 3$^\circ$ of the centre of the Sextans dSph (grey dots), consisting of 94\,770 objects that possess photometry, proper motions and parallax measurements. The horizontal and vertical axes are the local tangent plane coordinates corresponding to right ascension and declination, respectively. The positions of the 41 VLT/FLAMES LR8 pointings ($\sim25^\prime$ diameter) are shown with blue dotted and red full circles. In blue are the 16 previously analysed fields, and in red are the 25 new fields presented here for the first time (with field IDs, as listed in Table~\ref{table1}.). The nominal tidal radius of Sextans from \cite{Irwin95} is shown as an orange ellipse. 
}
\label{spatt}
\end{figure*}

\section{Introduction}

The Sextans dwarf spheroidal (dSph) galaxy is the faintest and most diffuse of the so-called {\it classical} dSph around the Milky Way. It was found in 1990, among the last Local Group galaxies to be discovered exclusively on photographic plates \citep{Irwin90}. The Sextans dSph was also one of the first resolved galaxies to be found by an automated scanning algorithm, picking out an enhancement of stellar point sources despite being barely visible on the plate to the human eye. Sextans remains a difficult galaxy to display without resorting to image processing tricks. This is partly caused by its relatively low Galactic latitude ({\it l}$=243.5$ {\it b}$=+42.3$), which results in significant contamination by foreground Milky Way disk stars. Most critically, the Sextans dSph has a very low central surface brightness ($\Sigma_V \sim 26.2$ mag/arcsec$^2$) and a large physical extent on the sky with very few member stars per arcsec$^2$. 

The distance to the Sextans dSph has been determined in a number of different ways making use of the stellar population in the central region. The most recent study of the variable star population \citep{Vivas19} has detected 199 RR~Lyrae and 16 dwarf Cepheid variable stars. Of these, there are 41 RR~Lyrae with complete coverage of their light curves, and using the Period-Luminosity-Metallicity (PLZ) relation \citep[][]{Sesar17}, the distance is found to be 84.7~kpc. This agrees well with previous distance estimates \citep[e.g.][]{Irwin95, Mateo95,  Lee09, Medina18}.

The structural properties, including the precise centre of the galaxy, have always been challenging to define as they require wide-field sensitive and uniform imaging to improve on the early photographic plate results \citep{Irwin95}. The structural parameters vary quite a bit between different studies  \citep[e.g.][]{Irwin95, Roderick16, Okamoto17, Cicuendez18, Tokiwa23}. For example the nominal King tidal radius has been determined with a range from $160^\prime \pm 50$ to $ 82^\prime \pm 7$, which means from 4 to 3~kpc at the distance of the Sextans dSph. \cite{Cicuendez18} provide a detailed overview of the complex situation and suggest a { nominal} King tidal radius of $120^\prime$.
 { The variety of values found in the literature is due to the difficulties in treating the extremely low surface brightness outer regions of Sextans, where few individual stars can be unambiguously identified as members on the basis of photometry alone.} 
Radial velocities from stellar spectroscopy and proper motions of individual stars provide uniquely valuable information to disentangle the sparse Sextans stars from those of the Milky Way and hence to clean up our picture of the structural properties of the Sextans dSph. 

\begin{figure*}[t]
\centering
\includegraphics[width=1\linewidth]{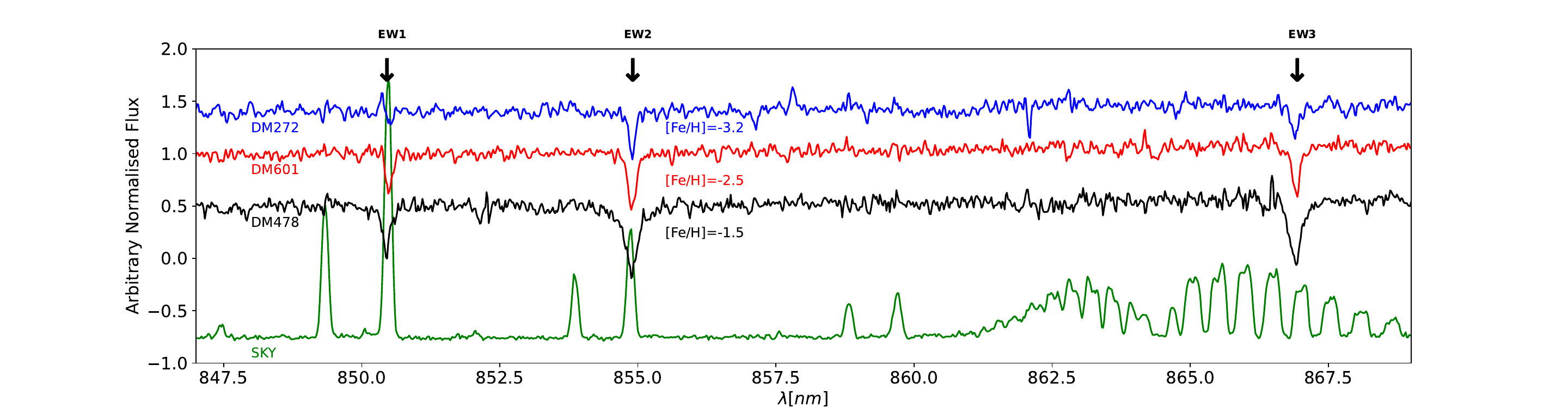}
\caption{Sky-subtracted and normalised VLT/FLAMES spectra of 3 Red Giant Branch stars in the Sextans dSph (DM272, DM601 \& DM478), with a range of \feh $\sim -1.5$ to $-3.2$ and $\vlos\sim 230$--$240$~km/s and all with $\text{S/N}\sim 35$ { per pixel}. The three Ca~II triplet absorption lines are easily visible in each spectrum at $\lambda\lambda$8505\,\AA (EW1), 8550\,\AA (EW2), 8670\,\AA (EW3) indicated with black arrows at top of the plot. The green sky spectrum is also normalised and plotted on the same scale. All spectra except DM601 have been arbitrarily shifted in flux to allow the spectra to be plotted together.}
\label{spec}
\end{figure*}

A pre-\gaia\ ground-based proper motion study, using Subaru Suprime Cam images taken 10 years apart \citep{CD18}, obtained a global proper motion of ($\mu_\alpha \cos\delta$, $\mu_\delta $) = ($-0.409 \pm 0.050, -0.047 \pm 0.058$) mas yr$^{-1}$. This is 
within the errors of the \gaia\ eDR3 proper motion 
($\mu_\alpha \cos\delta$, $\mu_\delta $) = ($-0.40 \pm 0.01, 0.02 \pm 0.02$) mas yr$^{-1}$
\citep{Battaglia22}. Numerous \gaia\ based determinations of the membership probabilities of individual stars in the Sextans dSph have been made, mostly to determine the dynamical motion of Sextans around the Milky Way, but also to better define its structural properties \citep[e.g.][]{PaceLi19, McV20, Martinez21, LiHammer21, Pace22, Battaglia22, Jensen24}. 

There have also been a number of low- and high- resolution spectroscopic studies of increasingly large samples of stars in the Sextans dSph \citep[e.g.][]{DaCosta91,Suntzeff93, Hargreaves94, Kleyna04,  Walker09, Kirby10, Battaglia11, Theler20, Walker23}. These have shown that Sextans is a predominantly metal-poor (MP) galaxy. The presence or absence of cold substructures and kinematic anomalies has been debated back and forth over the years \citep{Kleyna04, Walker06, Battaglia11, CicBat18, Kim19}. 
Several detailed chemical abundance analyses have been made, often with particular attention paid to the search for the most MP stars \citep[e.g.][]{Shetrone01, Aoki09, Tafelmeyer10, Theler20, AokiM20, Mash22, Roederer23}. The scatter of the $\alpha$-element abundances compared with [Fe/H] in the Sextans dSph { in some of the older studies} have been  { found} to be larger than for other dSph galaxies \citep[e.g.][]{Aoki09}. This maybe due to early chemical inhomogeneities, which are expected { to be large} in very low-mass dwarf galaxies \citep[e.g.][]{Roederer23}. However, { more recent} studies have { found} less scatter { than the older studies} \citep[e.g.][]{Lucchesi20, Theler20}. { This suggests that the question of the amount of scatter of abundances in Sextans stars requires more study.}

There have been several Colour-Magnitude Diagram (CMD) analyses of the Sextans dSph, although these are challenging outside of the central region because of the large extent of the galaxy on the sky and the difficulties in assigning membership of individual stars to Sextans based solely on photometry. The inner regions are well studied and there have also been impressive forays into the outer regions \citep[e.g.][]{Bellazzini01, Lee03, Lee09, Okamoto17, Cicuendez18, Bettinelli19}. These results are consistent with spectroscopic studies showing that Sextans has a predominantly old and MP stellar population, but there is also a more metal-rich (MR) component and there appears to be an age gradient, where the central regions are slightly younger than the outer regions \citep[e.g.][]{Okamoto17}.

As part of this VLT/FLAMES spectroscopic survey we also include a sample of stars in the distant halo globular cluster Pal~3 due to its intriguing and well known proximity to the Sextans dSph. This cluster was discovered well before it was realised that it was close to a low surface brightness galaxy \citep{Abell55}, and has at some points in the past been erroneously thought to be a galaxy, given its extreme isolation and distance from the Milky Way for a globular cluster. Pal~3 has previous spectroscopic measurements of radial velocities and abundance analysis \citep[e.g.][]{Peterson85, Koch09}, as well as proper motion determinations with \gdr{3} \citep[e.g.][]{VasiBaum21, BaumVasi21}, and detailed photometric studies, most precisely with the Hubble Space Telescope  \citep{Stetson99}.
Its orbital properties clearly indicate an accreted origin, although its progenitor is still debated \citep[e.g.][]{Massari19, Callingham22}.

This paper brings together all the available VLT/FLAMES low-resolution spectra covering the Ca~II triplet (CaT) wavelength range (850--870nm) using the LR8 grating. This provides the most extensive overview of the global properties of the resolved stellar population of this galaxy thanks to the accurate membership determination from spectroscopic radial velocities (\vlos) combined with \gdr{3} photometry and proper motion measurements.

\section{The VLT/FLAMES spectroscopic observations of individual stars in the Sextans dSph}

Here we present our new measurements of \vlos\ and [Fe/H] for new and archival data for individual stars in the Sextans dSph from the ESO (European Southern Observatory) Very Large Telescope (VLT) FLAMES/GIRAFFE spectra of RGB stars. These data were collected between 2003 and 2019 in two main observing campaigns\footnote{171.B-0588 PI: Tolstoy (DART LP); 0102.B-0786  PI: Tolstoy}, see Table~\ref{table1}. The LR8 grating was used, which covers the wavelength range between 8204\AA\ and 9400\AA\ at a resolving power, R$\sim 6 500$. The wavelength region contains the CaT absorption lines, which are used to obtain radial velocities and determine metallicities \citep[e.g.][]{Ad91, Rutledge97, Cole00, Tolstoy01, Battaglia08a, Starkenburg10}. The early observations consist of 16 FLAMES pointings and have been published by \citet{Battaglia11}. 
Here we re-analyse these archival observations and combine them with new observations of 25 additional pointings, mostly from 2019. This leads to more measurements in the outer regions of the Sextans dSph  (see Fig.~\ref{spatt}) and all measurements have been determined in a uniform way. We also analysed the reliability limits of the spectroscopic determinations of \vlos\ and [Fe/H] as apply to the particular challenges of the Sextans FLAMES data set.

\subsection{Data Analysis}

The Sextans dSph has some clear global differences compared to the Sculptor dSph data analysed in \cite{Tolstoy23}, hereafter, Paper~I. Sextans is much lower luminosity than Sculptor and is larger on the sky, with a much lower central surface brightness \citep{Mateo98} and so the number of members is a much smaller fraction of the background, and a smaller number in total, despite a similar number of pointings. The Sextans spectra have on average a higher Signal-to-Noise ratio (S/N) than those of Sculptor in Paper~I. The systemic velocity of Sextans is about 100km/s higher than that of Sculptor and this means that all three CaT lines lie directly under strong sky lines for Sextans stars. In Fig.~\ref{spec} we show examples of the spectra of 3 similar luminosity RGB stars with a [Fe/H] variation from $-1.5$ to $-3.2$ and a sky spectrum to show where the sky lines fall. Each of the RGB spectra have been divided by the flux in the continuum (factor $\sim 200-250$). For the sky spectrum, to put it on the same scale as the RGB spectra, the height of the sky lines were matched to the un-sky-subtracted but normalised spectrum of DM601, which required division by a factor of 620. All the spectra in Fig.~\ref{spec}, including the sky spectrum, come from different FLAMES pointings. 

\begin{figure}[ht]
\centering
\includegraphics[width=0.9\linewidth]{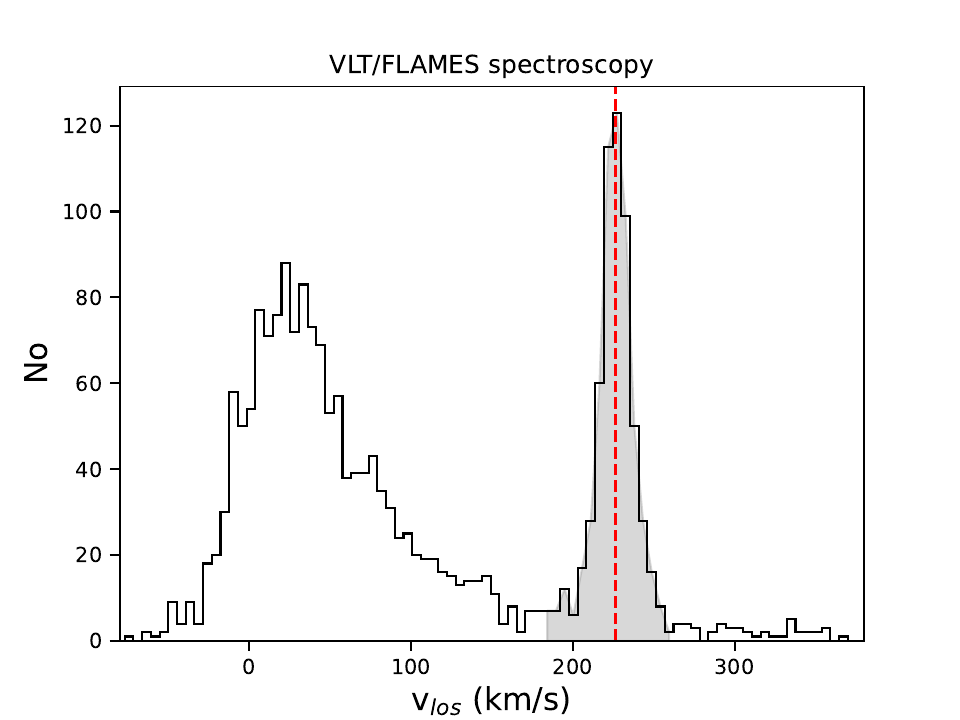}
\caption{Velocity histogram of the 2108 individual VLT/FLAMES spectra. The approximate range for members is shaded in grey ($\vlos=180$--$260$~km/s), with the expected mean velocity of members of the Sextans dSph  ($\vlos= 227$~km/s) as a red dashed line.} 
\label{velhist}
\end{figure}

We carried out the VLT/FLAMES spectroscopic analysis using the techniques as described in Paper~I. The first step in producing a uniform catalogue of \vlos\ and \feh\ measurements for RGB stars in the Sextans dSph is to process the 41 VLT/FLAMES LR8 pointings available in the ESO archive with the most recent ESO pipeline (via the esoreflex tool, ~\citealt{Freudling13}). This results in 2108 individual spectra, of which 989 are new. There are large numbers of non-member stars (see Fig.~\ref {velhist}) observed due to the  difficulties in separating members and non-members based purely on photometric colours, in pre-\gaia\ spectroscopic selections.
Our initial sample of likely members includes 388 spectra having \vlos\ in the range $180-260$ km/s. Of these 269 have been observed once and 58 have been observed twice and 1 star has been observed 3 times. The multiple measurements of the same star are combined to make a single averaged measurement. The total numbers of spectra and spectroscopic members per FLAMES field are given in Table~\ref{table1}. We eye-balled a number of individual spectra to determine how well the pipeline was working, paying particular attention to outliers and low S/N spectra as well as spectra with small EWs of the CaT lines.

As described in Paper~I, the ESO pipeline output spectra are sky subtracted and then the \vlos\ and the Equivalent Widths [EWs] of the CaT lines are determined using Mike Irwin's routines as described in Paper~I. Since Paper~I was published an upgrade of the sky-subtraction routine has been carried out (Mike Irwin, private communication), which is now more robust and better at identifying and removing sky subtraction residuals and other spurious noise. This was a critical improvement for the Sextans analysis as the radial velocity of the system means that relatively strong sky lines overlap all 3 CaT lines.
This can be seen by comparing the strong sky emission lines in the (green) sky spectrum in Fig.~\ref{spec}, compared to the weak absorption lines we are seeking to measure, especially at low metallicity stars found in Sextans. It can be seen that the most metal poor stars in ig.~\ref{spec}, the blue spectrum, at [Fe/H]$=-3.2$ of DM272, the weakest CaT (unused) line has a clear sky subtraction residual residual that dominates the line itself. Showing that 
 spurious residual noise from the sky subtraction a risk particularly at low S/N. We noticed that with the previous version of the pipeline the sky residuals could dominate also stronger the CaT lines and lead to inaccurate \vlos\ and [Fe/H] measurements. When the sky lines are so close to the CaT lines over-subtracting the sky can result in the measured \vlos\  being shifted towards that of the sky line, or even the sky residuals (which has zero velocity variation within the measurement errors) maybe mistaken for the CaT lines. This will tend to make the velocity dispersion tend towards zero if the effect is frequently present. These over-subtracted sky residuals can also lead to larger EWs than the true and thus a higher [Fe/H] than true. In contrast, the under-subtraction of the sky can lead to emission lines at the positions of the sky lines, but this will tend to be flagged as an obvious problem with the spectrum as emission lines are not expected in RGB spectra. We eye-balled spectra to make sure this kind of effect was reduced in the revised pipeline routines and that flagging spurious EW ratios or velocity differences between measurements using different techniques removed the remaining problem cases. With this new pipeline and careful flagging of problem spectra using the expected values of EWs and the repeatability of velocity measurements, the sky subtraction works well, even for low metallicity stars.  

As described in Paper~I, partly to increase the accuracy of the measurements and partly as a way to check that the spectrum behaves as would be expected for an RGB star, the \vlos\ is determined using three different methods (cross-correlation, line fitting and maximum likelihood) and [Fe/H] is obtained in two different ways (summing the area in the absorption lines and fitting the line) to determine the EWs of the two strongest CaT lines. These EWs are then used to determine [Fe/H] using the calibration developed by \citet{Starkenburg10}. { The CaT indicator has been shown to be robust also for varying values of [Ca/H]. This was shown by comparing CaT measurements of RGB stars in Sculptor and Fornax dSph with direct high-resolution [Fe/H] and [Ca/H] measurements \citep{Battaglia08a, Starkenburg10}. The [Fe/H] obtained from the CaT and directly from individual Fe lines were well matched for a large range in [Ca/H], a better match than to [Ca/Fe].}
This scaling relation has been { further} tested for [Fe/H]~$\geq -4$ \citep{Starkenburg13} which is important regime for a study of Sextans.

\begin{figure}[ht]
\centering
\includegraphics[width=1.0\linewidth]{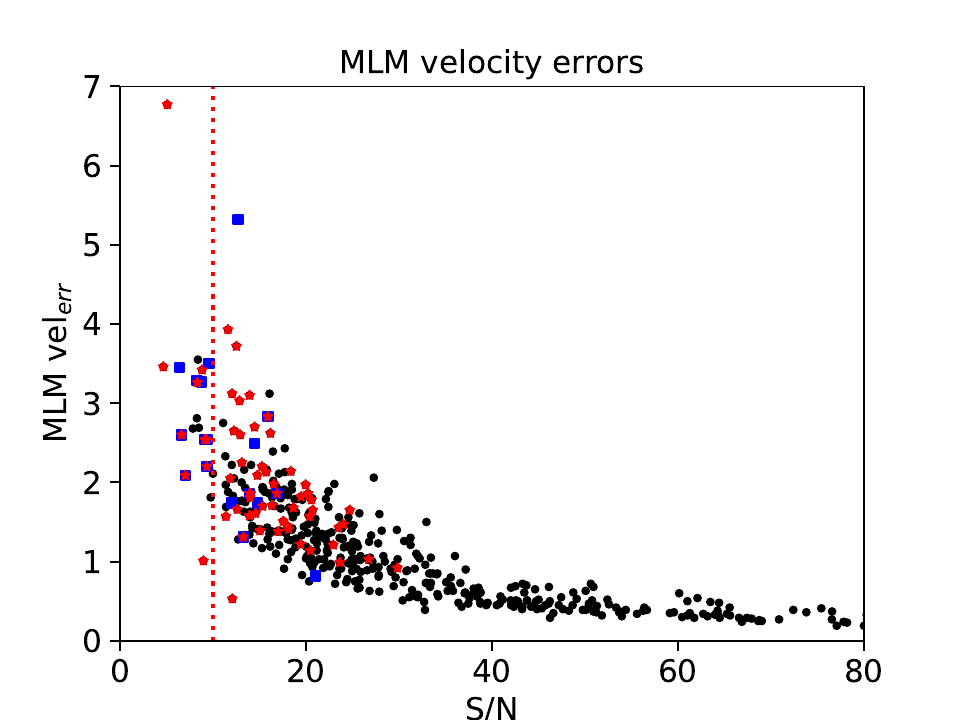}
\caption{The maximum likelihood method (MLM) velocity errors as a function of the signal-to-noise ratio (S/N) { per pixel} for measurements of \vlos\ for individual VLT/FLAMES LR8 spectra of likely RGB member stars in the Sextans dSph, with G$<19.7$. The blue squares are stars that don't pass the velocity quality criterion and the red star-symbols don't pass the EW quality criterion; some fail on both.
The vertical red dotted line indicates the acceptable limit (S/N$>10$) that is applied to the final selection of \vlos\ measurements, bearing in mind that the blue squares and red stars are removed before the cut. }
\label{errs2}
\end{figure}

\begin{figure}[ht]
\centering
\includegraphics[width=1.0\linewidth]{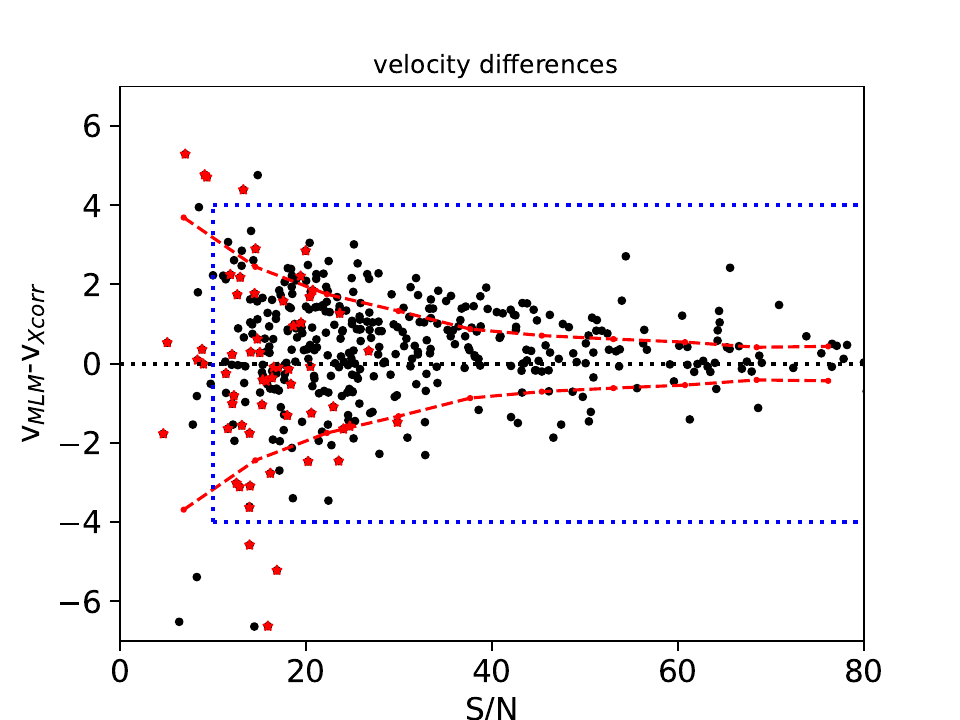}
\caption{Differences in \vlos\ as measured with the maximum likelihood (v$_\mathrm{MLM}$) and the cross-correlation (v$_\mathrm{Xcorr}$) methods as a function of the signal-to-noise ratio (S/N) { per pixel} for the individual VLT/FLAMES LR8 spectra of likely Red Giant Branch members of the Sextans dSph  with G$<19.7$. Red star-symbols are those stars that don't meet the EW criterion. The vertical blue dotted line shows the minimum acceptable $\rm S/N=10$, and horizontal blue dotted lines show the limits for outliers, at $\pm4$\,km/s. The black dotted line is at ${\rm v}_{MLM} -{\rm v}_{Xcorr} =0$. The red dashed lines are the running mean for the combined direct error estimates made on the individual velocity measurements from the two different methods.}
\label{sncut}
\end{figure}

\subsection{Quality Cuts}

From the measured \vlos\ from our FLAMES spectra there are clearly many obvious Galactic foreground stars in most pre-\gaia\ FLAMES fields (see Fig.~\ref{velhist}). In a few of the more recent outer fields the \gaia\ pre-selection produced such a small number of targets that experiments were made to include a wider selection of targets.
In addition, there are six stars in our spectroscopic sample that are likely members of the outer halo globular cluster Pal~3. 

Our exploration of quality cuts for the spectra are only carried out for stars where \vlos\ is found to be within the broad range expected for membership. A few of these may be rejected later when we look at their properties in \gdr{3}. { The S/N is determined by the pipeline from the continuum variation per pixel. There is a broad range of S/N in our measurements that does correlate with the magnitude of the star, with a degree of scatter coming from the variation of conditions under which data were taken over the years. The brightest stars, G$\sim 17$, typically have a S/N$\sim 100$, at G$\sim 18$ this falls to S/N$\sim 50$ and for G$> 19$mag, the maximum S/N is $\sim 30$.}

To understand the type of problems we find in our spectra of stars with \vlos\ consistent with membership in Sextans, especially at low S/N, we eye-ball a number of individual spectra. We also make check measurements that ought to be well behaved for RGB stars. Thus we investigate the difference in measuring \vlos\ from the same spectrum in independent ways, which we call the velocity difference (vdiff) criterion. This should be small for RGB spectra, as these are the type of objects that the pipeline is expecting. We also look at the ratio of the CaT EW2/EW3  measurements that are expected for RGB stars, which we call the EW criterion. This should be fixed at an fairly constant value for RGB stars. If EW2 and EW3 are contaminated by sky residuals, or cosmic ray residuals, then these should be different in the case of each line, and so the ratio will vary. Making ``quality cuts'' on the expected values for vdiff and EW criteria helps to remove most problems expected in the spectra above the S/N threshold. We can see from looking at the spectra that those with S/N$<10$ { per pixel} show a lot of scatter in the results caused by noise making the CaT lines difficult to measure, most especially at low metallicity. We prefer to make a cut that doesn't preferentially remove low metallicity stars, although the errors will always be higher for weaker lines as given by the error bars. In the next sub-section we will carefully check that the error bars are reliable. 

\subsubsection{\vlos\ quality cuts}
The errors on the \vlos\ measurements from the maximum likelihood method (MLM) method, over the entire survey (for S/N$<80$ { per pixel}), plotted in Fig.~\ref{errs2}, are well behaved. These formal errors, provided by the pipeline can be compared the velocity differences between two independent methods of measuring \vlos\ applied to the same spectrum. This is shown in Fig.~\ref{sncut}, where the MLM method (v$_{MLM}$) and the more straight forward cross-correlation method (v$_{Xcorr}$) are compared. In Fig.~\ref{sncut} the combined error estimates from the pipeline in Fig.~\ref{errs2}  are given as red dashed lines. These mean lines suggest that the  formal errors based on the ESO pipeline error arrays and the scatter from indivdual measurements are very similar, which means that the error determinations are reliable and well behaved down to our S/N threshold.
 There is a noticeable offset in the scatter in Fig.~\ref{sncut}, in that there are more positive velocity differences than negative, so v$_{MLM}$ tends to be slightly higher than v$_{Xcorr}$. This is likely due to the differences in the two methods of measuring \vlos . The MLM modelling has three fit parameters, overall scale factor, gaussian FWHM, and \vlos\ and makes use of the pipeline data error estimates. The X-corr method only considers \vlos\ and ignores everything else. 

Based on Fig.~\ref{errs2} and Fig.~\ref{sncut} we determine that the quality cut on the S/N of the spectra for reliable and accurate \vlos\  measurements is $\text{S/N>}10$ { per pixel}. We also determine, from visual inspection of a selection of spectra that stars with $| {\rm v}_{MLM} -{\rm v}_{Xcorr}| > 4$km/s  typically have some issue with either poor sky subtraction or perhaps spurious cosmic rays, or they are not an RGB star, so the stars outside our quality control selections in Figs.~\ref{errs2} and \ref{sncut} are verified to typically be spectra with problems, usually related to not being an RGB star.

\subsubsection{EW quality cuts}

\begin{figure}[ht]
\centering
\includegraphics[width=0.9\linewidth]{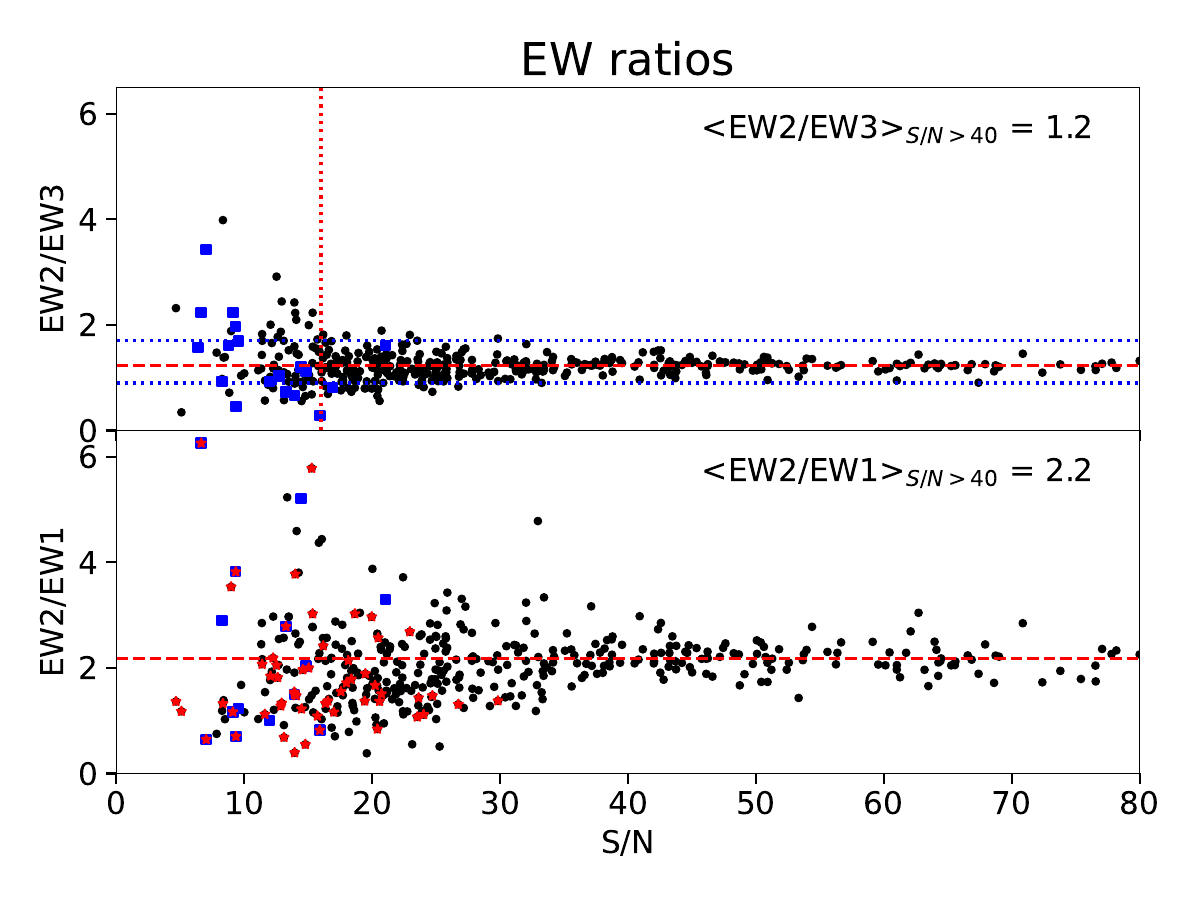}
\caption{Ratios of the EWs of the Ca~II Triplet lines EW3 and EW1 with respect to the strongest line EW2 
as function of the signal-to-noise ratio (S/N) { per pixel} for individual VLT/FLAMES LR8 spectra that are likely members of the Sextans dSph with G$<19.7$. Blue squares are spectra that don't meet the velocity difference criterion. 
The red dotted vertical line in the upper panel is at S/N~$= 16$ { per pixel}, which is the cut for accurate [Fe/H] measurements based on EW2/EW3.
The dotted blue horizontal lines are limits of the values of
the ratios at 0.9 < EW2/EW3 < 1.7. In the lower plot red star-symbols don't meet the limits in the upper plot for EW2/EW3. The red dashed horizontal lines in both panels show the mean of the ratios at a S/N$>40$ { per pixel}. }
\label{ews}
\end{figure}

\begin{figure}[ht]
\centering
\includegraphics[width=0.9\linewidth]{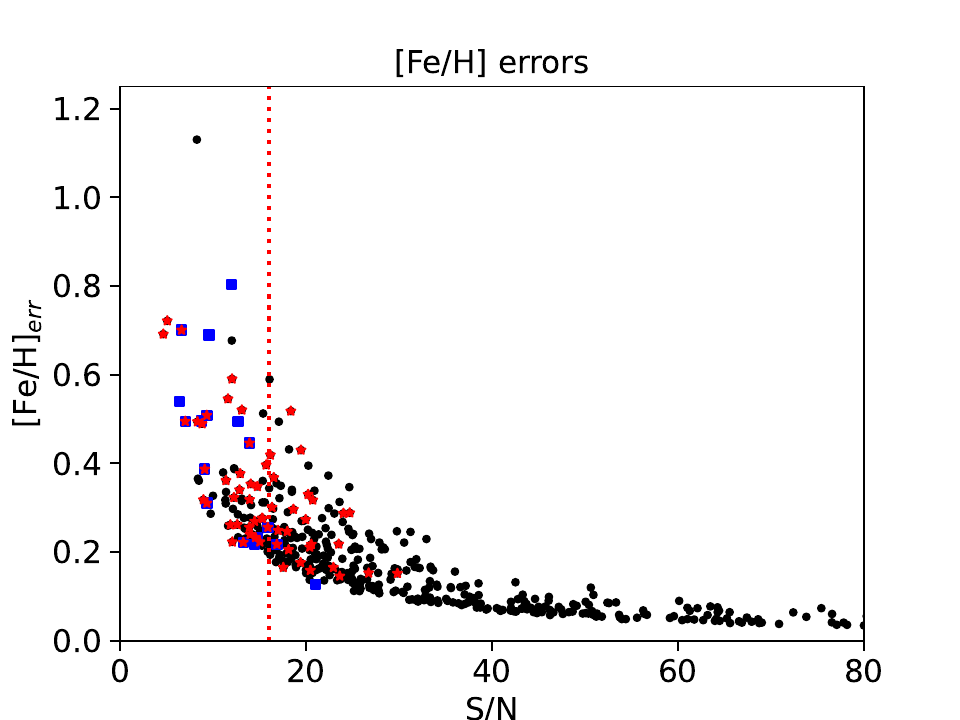}
\caption{
The [Fe/H] errors as a function of S/N { per pixel} for measurements of individual VLT/FLAMES LR8 spectra, consistent with membership in the Sextans dSph and with G$<19.7$. 
The blue squares are stars that don’t pass the velocity difference criterion and the red star-symbols don’t pass the EW criterion; some fail on both. 
The vertical red dotted line indicates the $\rm S/N=16$ { per pixel}, the cut that is applied to the final selection of reliable [Fe/H] measurements. 
}.
\label{sncut2}
\end{figure}

Accurate EW measurements of the CaT lines are more exacting than \vlos\ measurements, as EWs require a determination of the line shape as well as the peak. However, the ratios EW2/EW1 and EW2/EW3 will be around a fixed value for RGB stars. This is quantified in Fig.~\ref{ews}, where we show for the individual VLT/FLAMES LR8 spectra of \vlos\ likely members how these ratios behave. When the spread in these ratios increases for a given S/N, the measurements are less reliable because of increasing uncertainties. In the case of our spectroscopic sample this can be a useful indicator that either the object is not an RGB star or that the CaT lines are contaminated by sky residuals, both cases making the measurement unreliable and the object should therefore be removed from the sample. From Fig.~\ref{ews} it is obvious that the scatter is less for the ratio of stronger lines (EW2/EW3) and this is why these are the only lines we use in the CaT [Fe/H] determination  \citep[e.g.][]{Starkenburg10} and this is why it is also only ratio we use to select reliable spectra. This is our EW criterion. The
scatter in the ratio EW2/EW3 in Fig.~\ref{ews} appears well behaved for S/N~$\gtrsim 16$ { per pixel}. Therefore, this is our threshold for a reliable [Fe/H] determination. For lower S/N the errors become large enough to make the measurements too uncertain with a scatter that tends towards higher values of the ratio. This S/N threshold is higher than is required for accurate \vlos , as expected. This is also confirmed by looking at the [Fe/H] errors determined by the pipeline as a function of S/N { per pixel} (Fig.~\ref{sncut2}) and also from looking at individual spectra.

The EW2/EW3 ratio is not a perfect quality cut for our spectra, as if a spectroscopic target is not an RGB star in the Sextans dSph there can be no significant lines in the right places, and the pipeline can find similar noise at the positions of EW2 and EW3 resulting in a ratio around 1, consistent with the scatter around what is expected. Being able to tell the difference between a low metallicity RGB star with weak lines and a different type of spectrum, meaning not a member of Sextans, often requires visual inspection to check if a spectrum is clearly an RGB spectrum, with visible CaT lines. Also when we see clear signs of line blending in a spectrum, often leading to surprisingly high EWs, this suggests that the target is a double star (either intrinsic or by chance). Most objects like this were removed by the vdiff and EW quality cuts. One peculiar object remained (S07-80), with an unusually high [Fe/H]$=+0.16$. There are two spectra of this object, and both show possible signs of being two overlapping stars, one with narrow CaT lines and one with broad ones. We remove it from further consideration as a probable double spectrum.

\subsubsection{Making a final sample of FLAMES measurements}

It is clear that quality cuts depend on the use that will be made of the final set of measurements. To look for interesting and unusual populations for further follow-up it might be interesting to also consider observations at lower S/N, bearing in mind that there will be larger uncertainties. The cut for S/N$>10$ { per pixel} provides a uniform and well-defined spectroscopic data set to determine the global stellar properties of the Sextans dSph, which is the aim of this paper. 

Of the original 2108 spectra, 6 objects are not found in \gdr{3}, and so they are removed from further analysis, as would have happened anyway, as their \vlos\ are clearly inconsistent with being members. 

\begin{figure*}[t]
\centering
\includegraphics[width=1.0\linewidth]{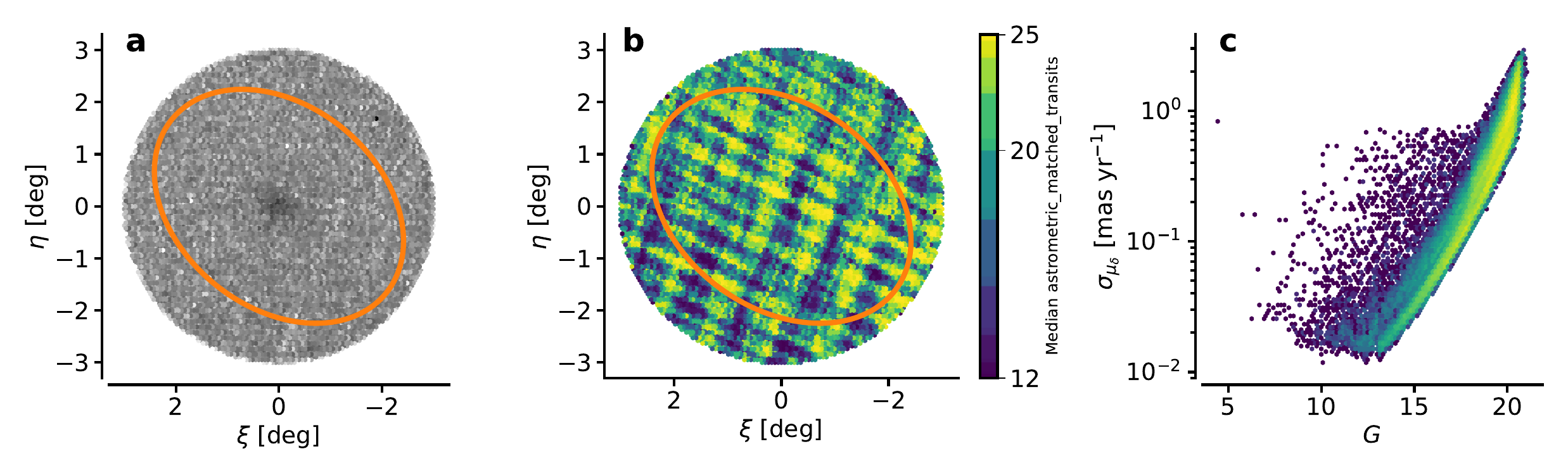}
\caption{\gdr{3} data within 3~degrees from the centre of the Sextans dSph.  a)~Distribution of sources on the sky; b)~Median number of astrometric observations (\texttt{astrometric\_matched\_transits}) matched to a given source; c)~Uncertainty on \pmdec\ as a function of $G$. The orange ellipse in panels a) and b) indicates the nominal tidal radius of the Sextans dSph.}
\label{gaia1}
\end{figure*}

\subsection{Large spectroscopic surveys with other telescopes}

Several other medium-resolution spectroscopic surveys of the Sextans dSph have been made with other telescopes. The most extensive survey, with publicly available tables, is the Michigan/MIKE Fiber Spectrograph (MMFS) survey \citep{Walker09}, recently updated and combined with a MMT/Hectochelle survey \citep{Walker23}. This study uses the Mg-triplet spectral range ($\sim 5150-5200$\AA ). It provides a huge database of measurements based on individual spectra, although many are of quite low S/N. There are 4257 Hectochelle spectra of stars in the direction of Sextans, and 2848 spectra from Magellan/M2FS, and an additional 213 M2FS spectra at medium resolution all matched to \gdr{3}. From these very large numbers we make a selection matching our own criteria, such that G$<19.7$ and $180<\vlos <260$km/s and we also require that the total number of observations of each source, after their quality-control filter flag has been applied, \texttt{ good\_n\_obs} is greater than zero. This results in 217 stars found in both surveys and 146 stars in the Walker et al. survey and not in FLAMES. 

The measurements in common between the FLAMES survey and the Walker et al. survey are used to make a comparison (see Appendix~B). There are 450 spectra of reasonable quality of the 217 stars in common, and these can all be used in the comparison. From this comparison we can see that the \vlos\ measurements agree well. 
Thus we can add the Walker \vlos\ measurements in our kinematic sample, after first determining the calibration offset between the two surveys, as also described in Appendix~B. Adding the Walker et al. measurements of \vlos\ for additional stars to our survey does not change our results. It fills in an area close to the centre of the Sextans dSph where FLAMES has not observed many targets and this is a region where structures have previously been found \citep[e.g.][]{CicBat18}. When we include the \vlos\ measurements from Walker et al. in our plots, 
we always use distinct symbols for this sub-sample. 

The [Fe/H] determinations between the surveys are also compared in Appendix~B, but in this case the scatter is quite large at any S/N and not symmetric around an offset value, so the [Fe/H] from \cite{Walker23} are not included in our plots or analysis.

New more complete samples will undoubtedly become available in the near future when DESI, WEAVE, 4MOST and the Subaru Prime Focus Spectrograph survey look at the Sextans dSph.

\begin{figure*}[t]
\centering
\includegraphics[width=0.9\linewidth]{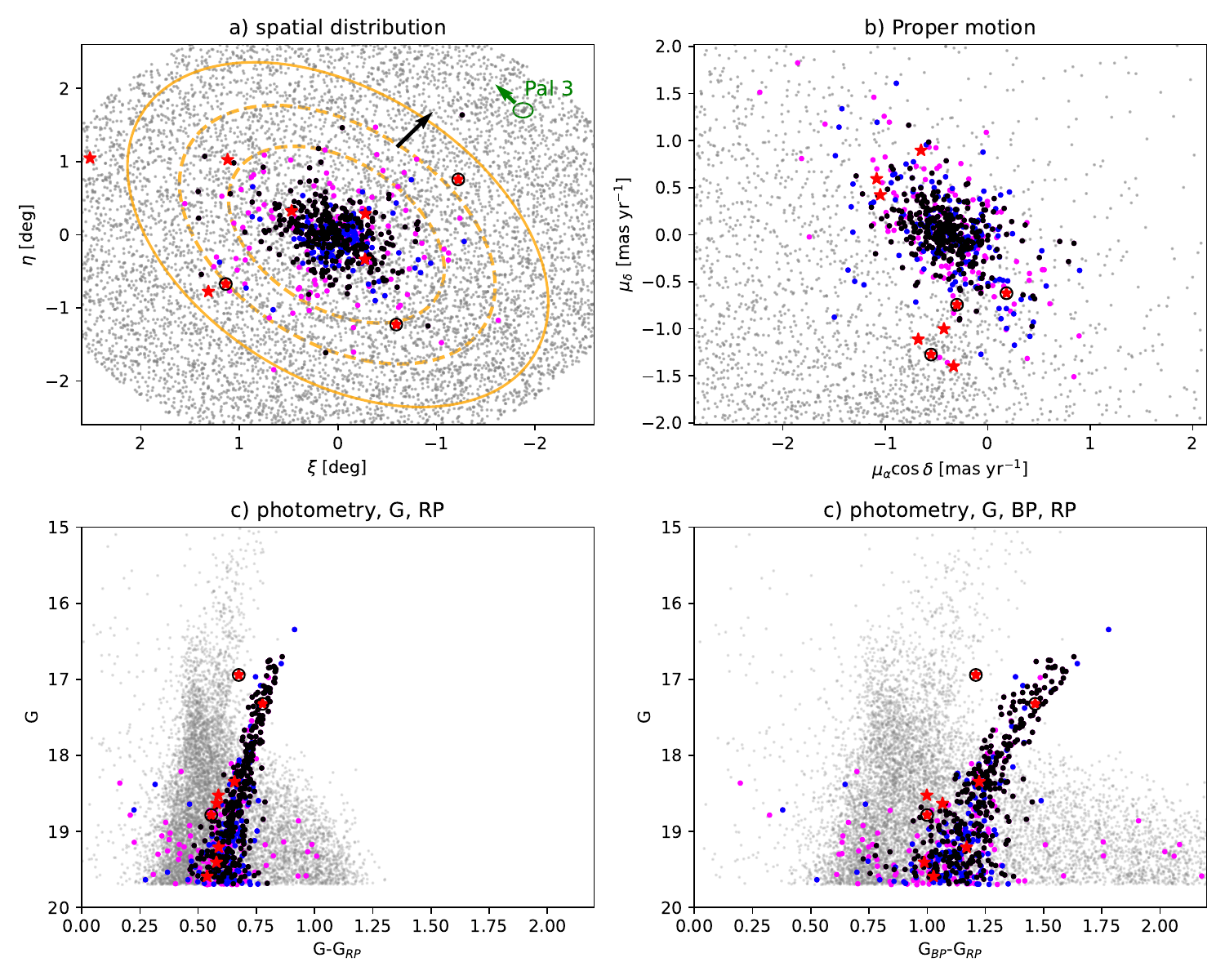}
\caption{\gdr{3}/FLAMES membership selection for the Sextans dSph: a) spatial distribution of \gdr{3} positions overlaid with the spectroscopic members. The largest solid orange ellipse is the nominal tidal radius from \cite{Irwin95} the dashed line ellipses are from \cite{Cicuendez18} (outer) and \cite{Roderick16} (inner). The black arrow is the mean proper motion on the sky of Sextans and the green arrow is the same for the Pal~3 globular cluster, which is at the position of the green ellipse; b) \gdr{3} proper motions in RA and Dec; c), d) \gdr{3} photometry as Colour-Magnitude Diagrams. In all plots, grey dots are the \gdr{3} catalogue entries, for $G<19.7$, the magenta dots are the 604 likely members of Sextans (P$_{mem} >0.07$). Over plotted as black dots are 333 \gaia/FLAMES members (with S/N$>10$ { per pixel} and $G<19.7$) and 146 blue dots from \cite{Walker23} that are not in the FLAMES catalogue. The 9 red stars are in the FLAMES sample with a \vlos\ consistent with membership, but just outside the \gaia\ membership criteria. The 3 red stars with an additional black circle have lower \vlos\ than the rest of the member stars (\vlos\ $< 190$ km/s), but still within 4$\sigma$ of the mean systemic \vlos .}
\label{sextmems}
\end{figure*}

\section{\gdr{3} aided membership determination}

For individual stars \gdr{3} provides parallax, proper motion in RA and Dec, as well as uniquely well-calibrated photometry, and these can all be combined with our spectroscopic radial velocities to provide a robust likelihood determination for stars being members of the Sextans dSph. Combining all this different information also means that we do not need to invoke too many prior assumptions in identifying members, which allows an unbiased assessment to be made of where the limits of membership lie. The exact approach used for Sculptor in Paper~I does not work for Sextans as the foreground (Galactic) contamination is much larger; the Galactic latitude is considerably lower, and the proper motions of Sextans are not as precisely measured nor are they significantly different from the foreground Galactic population. This  makes the inclusion of radial velocity measurements, coming from ground-based surveys all the more important.

 The exquisite photometry and astrometry from \gaia\ has motivated the development of Bayesian techniques that determine robust membership probabilities for individual stars in nearby dSph galaxies making use of detailed prior information that determine the expected properties of stars in a dSph galaxy compared to other stellar populations in the same field \citep[e.g.][]{PaceLi19, McV20, Battaglia22, Sestito23}. Different groups have focused on different aspects, and here we include spectroscopic radial velocities to refine the membership likelihood for a large sample of individual RGB stars found in the direction of the Sextans dSph, using the \cite{Battaglia22} approach.

 The Sextans dSph is a very low surface brightness system that is very large on the sky (see Fig.~\ref{gaia1}a) and thus it has always been challenging  to determine its global properties. There is a lot of foreground contamination from the Milky Way and it is also not well scanned by \gaia\ , as can be seen in the chequered pattern in the median number of astrometric measurements per source (see Fig~\ref{gaia1}b) and the scatter in the uncertainties in the proper motion measurements as a function of brightness (see Fig~\ref{gaia1}c). The number of \texttt{astrometric\_matched\_transits} varies from 9 to 28, with a median of 19. For Sculptor dSph (Paper~I) the number of transits went up to $60-80$. The \gdr{3} catalogue is thus slightly less deep than it was for Sculptor, going down to G~$=20.85$ compared to $21.20$ for Sculptor \citep[as measured by the \texttt{M10} parameter,][]{Cantat2023}.

For Sculptor (Paper~I) the stellar population of the galaxy clearly stands out in the CMD
without any processing, because it is a higher density system embedded in a lower density Milky Way foreground. Sculptor also 
stands out well in parallax and proper motion plots, and thus a simple $\chi^2$ 
membership selection based on the astrometry worked well.
For the Sextans dSph, the histogram of the \gdr{3} calculated membership score, z \citep[as explained in][]{Tolstoy23}, based on proper motion and distance measures and their errors, shows no structure { that makes clear where a membership cut should be made}.  A selection for z$<14.2$ does bring out the CMD of Sextans more clearly. However, there remains a lot of Milky Way contamination. Thus, we should also take into account the spatial and CMD distribution of the stars in the field to improve the selection of the likely Sextans members. This is already done with a 
mixture model in \cite{Battaglia22}, and so we use their methodology to update their membership probabilities on the wider area covered by our spectroscopic survey and using the \vlos\ spectroscopic information where possible (see Sect. 4.2.1 of their article).

\begin{figure}[ht]
\centering
\includegraphics[width=1\linewidth]{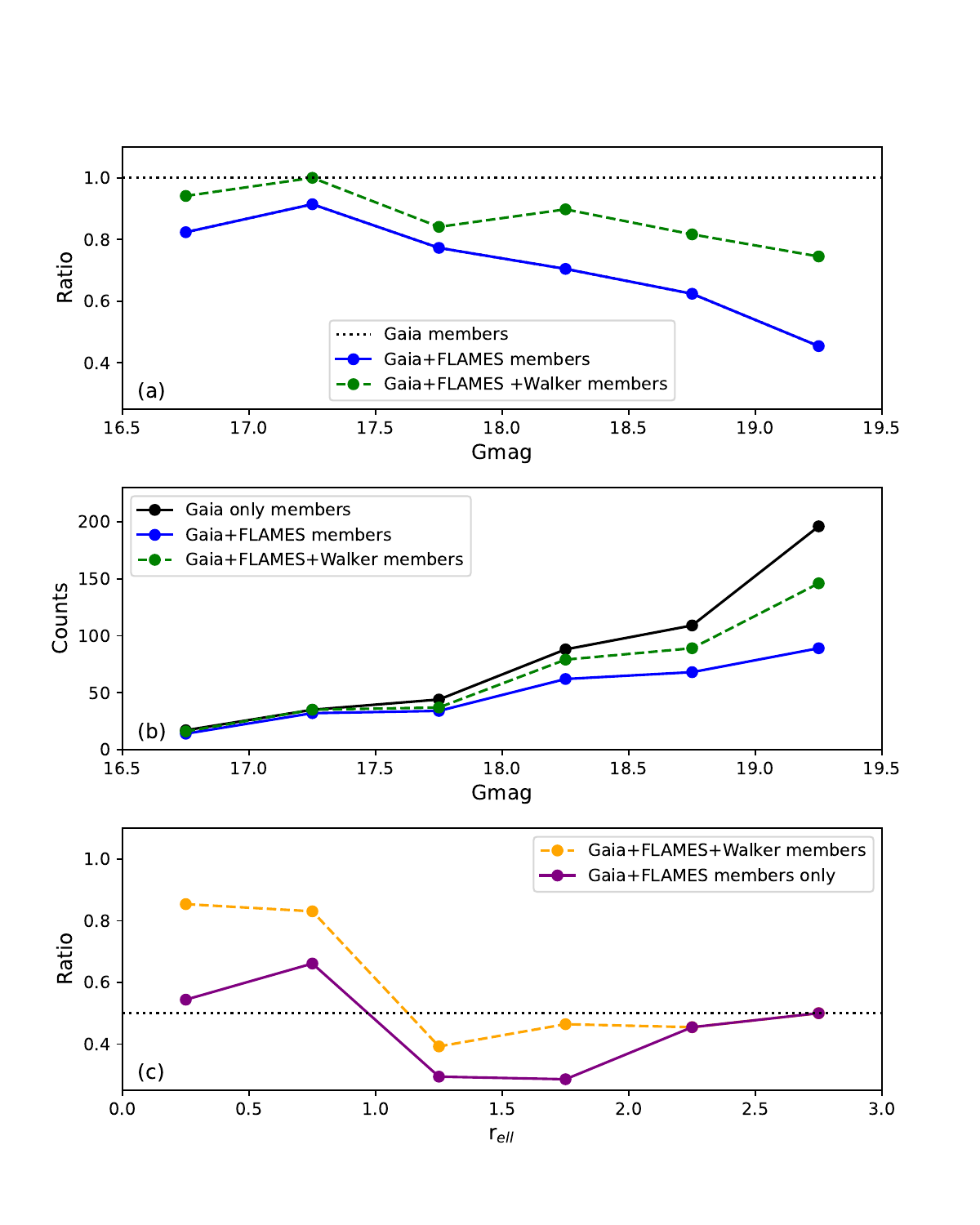}
\caption{The completeness of the VLT/FLAMES LR8 survey relative to the complete \gdr{3} members (G$<19.7$ and P$_{mem}>0.07$) for the Sextans dSph as a function of \gdr{3} G magnitude and Elliptical radius (r$_{ell}$), including the increased completeness from adding in  \cite{Walker23} \vlos\ measurements: a) as a fraction of the total \gdr{3} members (black), those with FLAMES confirming spectra (blue) and those with FLAMES plus Walker23 (green); b)
the total number of stars, with measurements from \gdr{3} (black), \gdr{3}+FLAMES (blue) and \gdr{3}+FLAMES+Walker (green); c) the completeness as a function of r$_{ell}$,  for \gdr{3}+FLAMES (purple) and \gdr{3}+FLAMES+Walker (orange), with a dotted black line at 50\% completeness }

\label{figcomp}
\end{figure}

From the \gdr{3} catalogue, we extracted all the sources with 3 colour photometry and astrometry (parallax and proper motions) within a circle of radius 3~degrees around the central position of the Sextans dSph \citep{Irwin95}, see Fig.~\ref{gaia1}a, which is 94~770 sources. By far the majority of these sources are not members of Sextans, and they have a membership probability of P$_{mem} = 0$
in \cite{Battaglia22}.

\gdr{3} 
information alone already presents an excellent  probability of membership (P$_{mem}$). The inclusion of the FLAMES \vlos\ information results in a more precise identification of highly likely members and obvious non-members and of course provide more information about each star.

We looked for a value of P$_{mem}$ below which our spectra were mostly clearly velocity non-members, and above mostly velocity members. This leads to a cut for likely membership where, P$_{mem} >0.07$ for stars with G$<19.7$. This leaves 604 likely members based on \gdr{3} probabilities alone. The magnitude limit, G~$<19.7$, is applied because the \gaia\ astrometry is less reliable for the fainter stars and this is also the case for the \vlos\ measurements from FLAMES, and so the membership probabilities are less reliable for fainter stars.

A FLAMES \vlos\ consistent with membership in the Sextans dSph is available for 333 \gdr{3} member stars,  with S/N~$>10$ { per pixel}. This is a little over half the likely members (604) found using only the \gdr{3} information for stars with G~$<19.7$. { The rest of the \gdr{3} likely members are not in our FLAMES sample.}
These are all in the expected range of  $\pm5 \sigma$ around the mean of the \vlos\ , showing the power of \gaia\ proper motions for membership selection. In addition, there are 9 stars that have P$_{mem} =0$, but \vlos\ in the expected range of membership. Looking at these stars more carefully, we see that they are at the edge of several selection criteria for \gdr{3} membership and they were removed because of this. They all have good \texttt{RUWE} values and a good number of visibility periods, and only three of the stars have an elevated crowding (\texttt{pd\_gof\_harmonic\_amplitude}~$\gtrsim0.1$), but these are the ones closest to the centre of Sextans. So there is no strong indication (also from looking at the spectra) that these stars are binary stars. Thus it is plausible that they could be members and they are retained as possible members, and plotted with different symbols in all future plots. 

An overview of the global picture from \gdr{3} is displayed in Fig.~\ref{sextmems}, where we show the collective distribution of the different properties we can measure for the resolved stellar population of Sextans dSph. We identify the different samples with different symbols in all plots.

\subsection{Completeness of the \gdr{3}-LR8 survey}

Fig.~\ref{sextmems}a shows very few stars outside the nominal tidal radius of the Sextans dSph as defined by \cite{Irwin95}. It also shows that there is a (small) hole in the spatial coverage of FLAMES measurements (black dots) in the central region.
There are velocity measurements from Walker at this position (blue circles) instead.
They do not significantly change the analysis, but they are identified in the plots as the scatter is often quite large.

In Fig.~\ref{figcomp} we show the overall completeness of the spectroscopic follow-up in the Sextans dSph. The \gdr{3} selection function \citep{Cantat2023} shows that in the sky area of  Sextans  \gdr{3} is complete to $G=20$. In Fig.~\ref{figcomp}a we show the fraction of the complete sample of \gdr{3} likely members that have FLAMES spectroscopic follow-up. Adding in the Walker et al. sample gives close to a 90\% completeness for the spectroscopic follow-up of stars with G~$<18.5$. It can be seen that the total number of stars rapidly increases with decreasing G in Fig.~\ref{figcomp}b, and the completeness with distance from the centre of the galaxy is decreasing, in Fig.~\ref{figcomp}c, although the total number of stars also decreases rapidly from the centre of the galaxy.  The spatial distribution of the \gdr{3} members can be seen in Fig.~\ref{sextmems}a. { The improved completeness in the outer regions comes from a relatively small number of stars and despite the effort that it is to find them, they are important to understand the true extent of the system and its properties and we have provided a more complete picture of Sextans than was previously available. }

\begin{figure*}[t]
\centering
\includegraphics[width=0.9\linewidth]{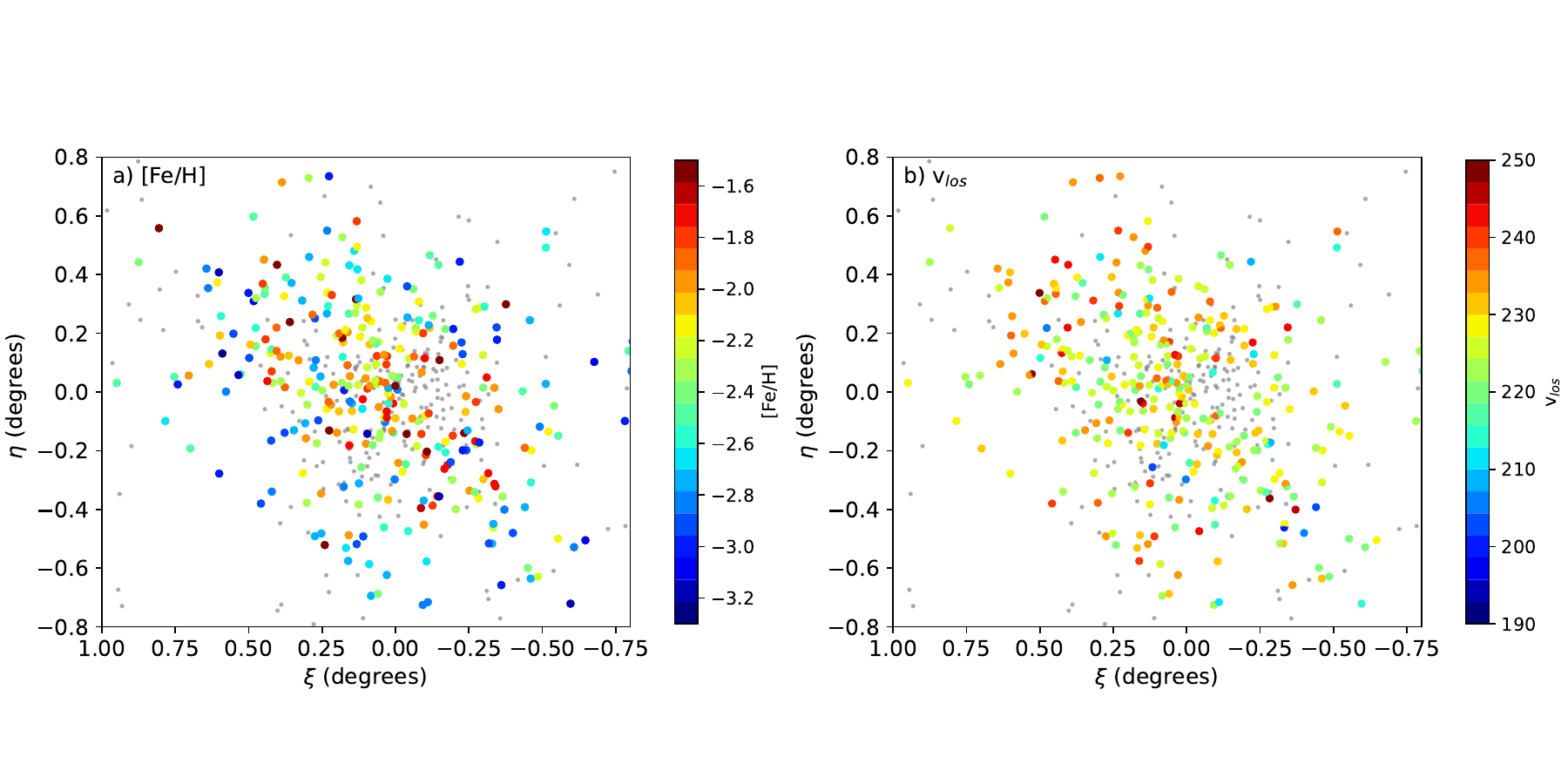}
\caption{ The spatial distribution of the \gdr{3}/FLAMES members across the Sextans dSph: a) coloured by [Fe/H];  b) coloured by \vlos . The grey dots are the \gdr{3} likely members without FLAMES spectra.
} 
\label{spatmem1feh}
\end{figure*}

\section{Results}

From the data analysis and quality criterion cuts in the previous section we have a catalogue of 604 likely members of the Sextans dSph based on \gdr{3}, and of these 325 have supporting \vlos\ measurements of acceptable quality from our FLAMES spectroscopic sample. There are an additional 
9  probable members based on their spectroscopic \vlos\ and borderline non-member \gaia\ parameters. We include these additional stars in our plots as distinct symbols, and they are not included in global averages or analyses.
Of these, 312 also have [Fe/H] measurements of acceptable quality, which includes these 9 probably members. 
An additional 146 of our 604 \gdr{3} members have \vlos\ measurements from \cite{Walker23}, these are included in the plots with different symbols. Making a total of 480 stars with \vlos\ spectroscopic measurements as well as \gaia\  astrometry consistent with membership. 

The full sample of measurements of the 604 likely members of the Sextans dSph are presented online (Table E.1).  Table E.1 includes, where it is available, FLAMES spectroscopic information, and for those stars not in the FLAMES sample but with a reliable Walker \vlos\ measurement \citep{Walker23}, these are also included in Table~E.1. In addition, the FLAMES spectroscopic and \gdr{3} properties of the 9 probable members that are \vlos\ members but have uncertain Sextans membership based on \gdr{3} are included in Table~E.1. 

In Table~E.2 we also provide the \gdr{3} and spectroscopic information for the stars for which we have FLAMES observations but either are very clear non-members on the basis of their spectroscopic \vlos\ or their spectroscopic measurements were not found to be reliable enough. If the unreliable spectroscopic members are likely \gaia\  members then this also noted in Table~E.1 and E.2.

For the \gaia /FLAMES sample with P$_{mem}>0.07$ (so excluding the 9 stars with borderline \gaia\ parameters), combined with a \vlos\ consistent with membership, 
the mean FLAMES spectrosopic and astrometric parameters for the entire Sextans dSph galaxy are given in Table~\ref{tablepm}. These agree with previous determinations, with the exception of the mean metallicity, $\langle$[Fe/H]$\rangle$, which has become significantly lower, due to the complex metallicity distribution function (MDF) of Sextans with a metallicity gradient declining in the outer parts, making a mean metallicity highly dependent on the size and spatial distribution of a sample. The differences in the astrometric parameters with the full \gdr{3} sample  \citep[given by][]{Battaglia22} are within the uncertainties.

{ In Fig.~\ref{spatmem1feh}, we show the spatial distribution of the \gdr{3}/FLAMES measurements for the central region of Sextans colour-coded by [Fe/H] in Fig.~\ref{spatmem1feh}a and by \vlos\ in Fig.~\ref{spatmem1feh}b}.  There do appear to be some  clumps in the spatial distribution, but this is a common effect of small number statistics and there is nothing that stands out has having a single metallicity or a single \vlos. { There are structures in the spatial distribution of likely members in Fig.~\ref{spatmem1feh}, but always containing a range of [Fe/H] and \vlos\ values so that nothing that stands out as different from the overall population}. Evidence has been presented for sub-structures that do show coherent properties in { \vlos\ and relative metallicity} in the central region of Sextans \citep[e.g. mostly recently by][]{CicBat18} but we miss a large fraction of the region used in this work. This is the ``hole'' near the centre of Sextans for which there are measurements from the \cite{Walker09} survey which was used by \cite{CicBat18}.

\begin{table}
\caption{Mean properties of the Sextans dSph based on the FLAMES spectroscopically confirmed sample (sp), where the errors are the standard errors on the mean. }             
\label{tablepm}      
\centering                       
\begin{adjustbox}{width=\columnwidth,center}
\renewcommand{\arraystretch}{1.3}
\begin{tabular}{l c c}          
\hline\hline                     
Property & ~~Mean & ~~Previous\\ 
\hline                           
    $\rm \langle\feh \rangle$ & ~~{ -2.37} $\pm$ { 0.025} & ~~-1.9 $^{[1]}$ \\
    $  \rm \langle \vlos \rangle$ (km/s)& ~~{ 227.1}$\pm $ 0.49~ & ~~226.6 $\pm$ 0.6 $^{[1]}$\\      
    $\langle\mu_\alpha\cos\delta\rangle_{sp}$ (mas/yr) & { -0.394}  $\pm$ { 0.017} &  $-0.40  \pm 0.01 ^{[2]}$ \\
    $\langle\mu_\delta\rangle_{sp}$  (mas/yr) & { 0.035} $\pm$ { 0.015}  &   $0.02 \pm 0.01 ^{[2]}$  \\
    $\langle\varpi \rangle_{sp}$  (mas) & { -0.007} $\pm$ { 0.015} &   $- 0.013 \pm 0.004 ^{[3]}$ \\
\hline                           
\end{tabular}
\end{adjustbox}
    \begin{tablenotes}
      \footnotesize
      \item $^{[1]}$ \citet{Battaglia11}, based on early FLAMES observations
      \item $^{[2]}$ \citet{Battaglia22}, based on \gaia\ eDR3
      \item $^{[3]}$ \citet{Helmi18}, based on \gdr{2}
    \end{tablenotes}
\end{table}

\begin{figure*}[t]
\centering
\includegraphics[width=0.85\textwidth]{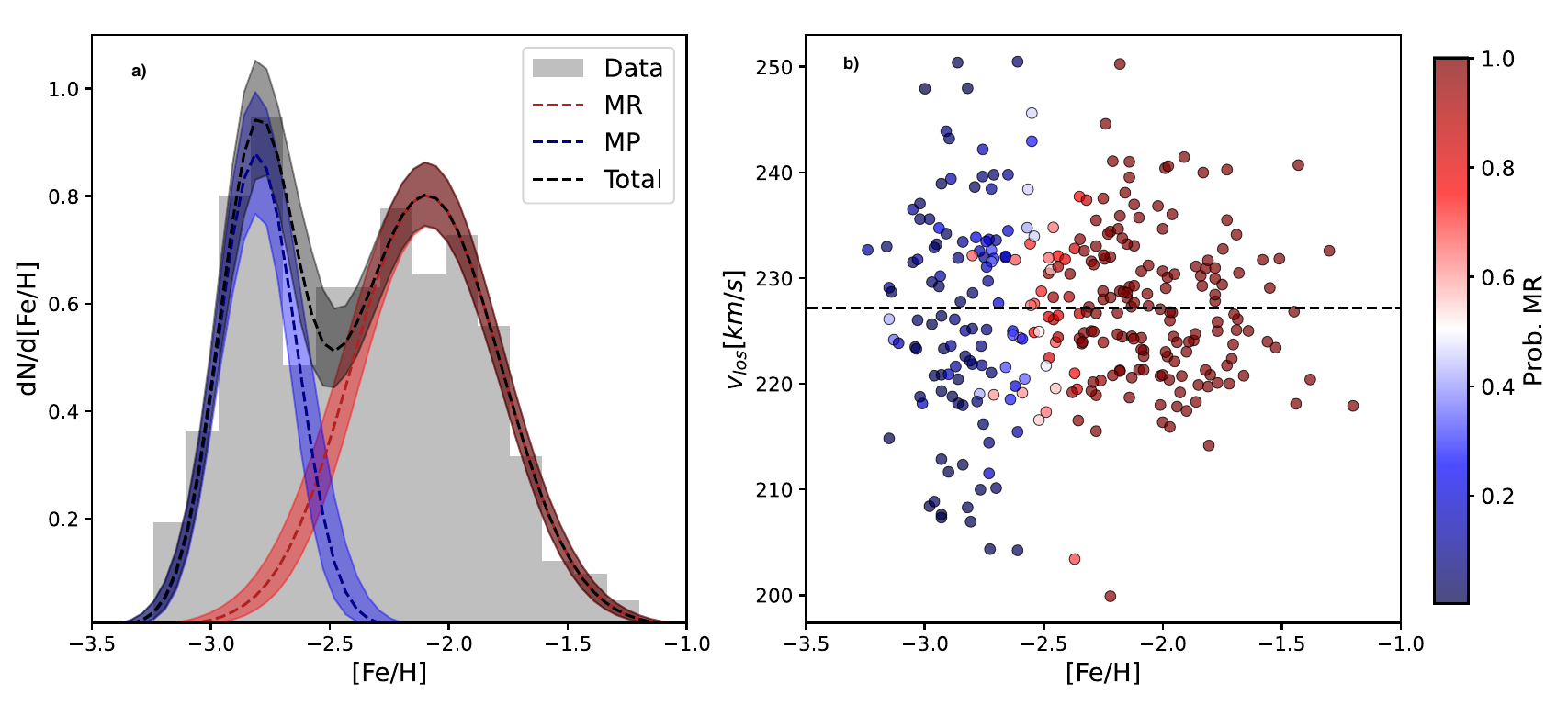}
\caption{Properties of the sample of spectroscopically observed member stars of the Sextans dSph, not including the 9 probable members, with reliable [Fe/H] information. Panel a): Normalized [Fe/H] distribution with the best-fitting 2-populations model. The dashed-lines indicate the predicted distribution of our model: red for the MR component, blue for the MP, and black for the sum of the two components as indicated in the legend. Bands indicate the 1$\sigma$ confidence intervals. Panel b): Projection of the stars in the \vlos\ vs [Fe/H] plane; the systemic velocity of Sextans is indicated with the horizontal line. The velocities have been corrected for the reflex motion between the Sun and Sextans. }
\label{fig:MR_MP_components}
\end{figure*}

\begin{figure*}[t]
\centering
\includegraphics[width=0.9\linewidth]{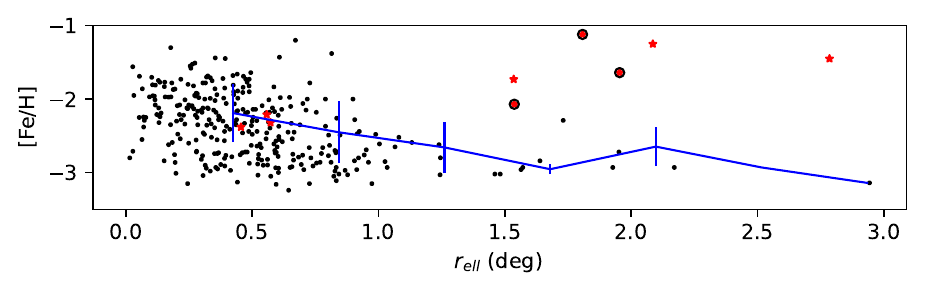}
\caption{The \feh\ for stars plotted with elliptical radius, r$_{ell}$, which is the projected semi-major axis radius, as black circles for the 312 measurements with S/N$>16$ { per pixel} in our Sextans dSph sample.   The blue line shows the binned mean and the error bars are the dispersion in each bin for the black points. The red star symbols are the 9 possible members of Sextans not included in the blue mean determination, with the 3 with unusually low \vlos\ circles in black.
}
\label{fehel}
\end{figure*}

\begin{figure}[ht]
\centering
\includegraphics[width=0.9\linewidth]{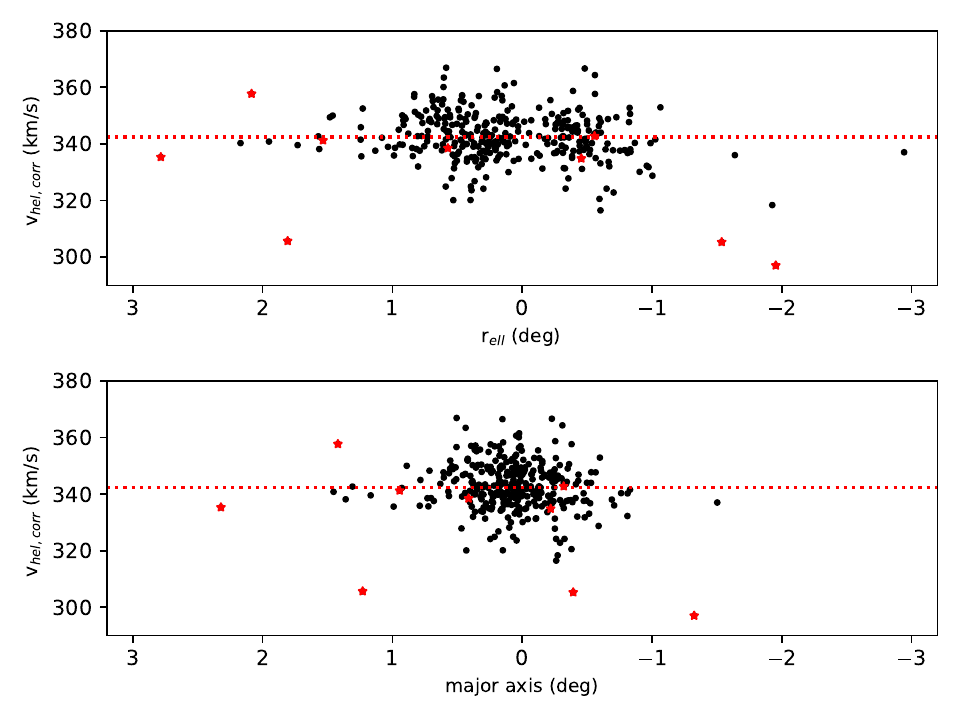}
\caption{VLT/FLAMES LR \vlos\ for individual members in the Sextans dSph, corrected for the shift created by the motion of the Sun around the Galactic centre at the position of Sextans, r$_{ell}$ (above); and the position along the major axis (below). 
The 9 probable members are shown as red stars. The mean corrected \vlos\ is a dotted line at 342.4 km/s.}
\label{fehel2}
\end{figure}

\subsection{The VLT/FLAMES spectroscopic metallicities}

In Fig.~\ref{fig:MR_MP_components}a we show the histogram distribution of the FLAMES [Fe/H] measurements, with S/N$>16$ { per pixel} and G$<19.7$ for RGB stars in the 
Sextans dSph. It clearly has a a double-peaked MDF, with a MR and MP peak. The comparison of the kinematic properties of the MR and the MP  population is shown in Fig.~\ref{fig:MR_MP_components}b where it can be seen that the MP population has a higher velocity dispersion than the MR (see also Table~\ref{tab:Chemodynfit}). The colour code in Fig.~\ref{fig:MR_MP_components}b shows the probability of a star belonging to the MR population, showing the clear distinction between the bulk of the two populations.

\begin{table*}[]
    \centering
    \caption{Best-fit parameters for the global properties of the metal-rich (MR) \& metal-poor (MP) chemo-kinematic stellar components of the Sextans dSph. }
    \label{tabbestfit}
    \begin{tabular}{|c|c|c|c|c|c|c|}
    \hline
         & $f$ & $\mathcal{M}$ & $\sigma_{\mathcal{M}}$& $\mathcal{V}$ [\kms]&$\sigma_{\mathcal{V}}$ [\kms] &$R_h[deg]$\\
        \hline
        MR & $0.63^{+0.05}_{-0.05}$ & $-2.09^{+0.04}_{-0.04}$ & $0.30^{+0.023}_{-0.022}$ & $227.3^{+0.5}_{-0.5}$ & $7.0^{+0.4}_{-0.4}$ &  $0.339^{+0.021}_{-0.021}$\\
        MP & $0.37^{+0.05}_{-0.05}$ & $-2.81^{+0.03}_{-0.03}$ & $0.13^{+0.03}_{-0.03}$ & $226.3^{+1.0}_{-1.0}$ & $10.5^{+0.8}_{-0.7}$ &  $0.62^{+0.05}_{-0.05}$    \\
        \hline
    \end{tabular}
    \tablefoot{$f$ is the fraction of the total number of stars belonging to the population; $\mathcal{M}$ \& $\sigma_{\mathcal{M}}$ are the mean and dispersion of the metallicities in the component; $\mathcal{V}$ [\kms] \& $\sigma_{\mathcal{V}}$  are the mean and dispersion of the \vlos\ in the component; $R_h[deg]$  is the projected half-light elliptical radius of the Plummer number density profile. }
    \label{tab:Chemodynfit}
\end{table*}

In Fig.~\ref{fehel} we show the distribution of the FLAMES [Fe/H]
measurements as a function of elliptical radius (r$_{ell}$), showing a clear decline of the mean [Fe/H] with increasing distance from the centre of the galaxy albeit with a large scatter. The 9 stars with borderline \gdr{3} membership properties present an intriguing subset. Of this sample, 6 are at r$_{ell} >1.5$~deg from the centre. There are only 8 other members in this extremely low surface brightness outer region of this galaxy. It is difficult to interpret the fact that the mean [Fe/H] of these 6 outer probable-member stars are clearly $\gtrsim1$~dex more metal rich than all the other 8 stars in these outer regions. There is no particular distinction in the spatial distribution of these probably members (see Fig.~\ref{sextmems}). It would be easy to interpret this as a sign that these stars must be non-members. But we think it must be left open for now. There are 3 additional stars with borderline \gaia\ membership at around r$_{ell} \approx 0.5$~deg and they are more MP than the outer 6, and they fit in well with the rest of the stars found around this distance from the centre of the galaxy. These 3 are clearly less contentious, but shows the risks of deciding membership too strongly on the basis of expectations of the results. For now this must remain an unsolved  conumdrum.

\subsection{Global kinematic properties}

In order to quantify the presence or absence of a line of sight (l.o.s.) velocity gradient, we follow the approach of \cite{Taibi18} and compare two different models: one exclusively supported by velocity dispersion and one allowing for the presence of a deviation from the systemic velocity that is a linear function of the distance from the Galactic centre. We first correct the l.o.s. velocities for the relative motion between the Sextans dSph and the Sun, hence removing perspective l.o.s. velocity gradients.
If we consider only the VLT/FLAMES data set, we obtain a systemic velocity of 227.0 $\pm$ 0.5 km s$^{-1}$ and a l.o.s. velocity dispersion of 8.3$_{-0.3}^{+0.3}$km s$^{-1}$ in both models. The l.o.s. velocity gradient is consistent with zero within little more than 1-$\sigma$ (2.0$_{-1.4}^{+1.4}$km s$^{-1}$ deg$^{-1}$) and given that in our Bayesian analysis, the model without rotation is moderately favoured with respect to that with rotation (by 2.1 points, using the Jeffrey's scale), we conclude that there is no evidence of a l.o.s. velocity gradient in this new VLT/FLAMES data-set. The results do not change if we consider the VLT/FLAMES data with the addition of the \cite{Walker23} measurements (the l.o.s. velocity gradient in this case is $2.2\pm1.3$ km/s/deg, fully consistent with the previous value).  

There is no sign of any structure in the \vlos\ corrected for the Galactic standard of rest plotted against position along the major axis and as a function of elliptical radius (r$_{ell}$) in Fig.~\ref{fehel2}.  The number of stars in the outer regions is relatively small, so it is hard to be conclusive about how or if the properties change towards the outer regions. The 6 outer probable-members (plotted as red stars in Fig.~\ref{fehel2}) look a better match to the more certain measurements than is seen in Fig.~\ref{fehel}, except for the 3 stars that  have unusually low \vlos\ which are red stars circled in black in  Fig.~\ref{fehel}. If these 6 stars are ignored then the velocity dispersion appears to be quite a bit lower in the outer regions compared to the central regions. This is less clear if the probable members are included. But in either case the numbers are very small.

There are known to be intriguing features in the spatial distribution of kinematic data that have recently been explored \citep[e.g.][]{Roderick16, CicBat18}. For example, \cite{CicBat18} found kinematic anomalies in the Sextans dSph, in the form of a ``ring-like'' feature, exhibiting a larger mean velocity and lower relative metallicity than the control sample. This feature was found only in the spectroscopic data-set of \cite{Walker09} because the spatial coverage of the other large spectroscopic surveys existing at the time \citep{Kirby10,Battaglia11} did not  sufficiently cover the region where the feature is more evident. As the new data set presented here does not add significant coverage in the region of the ``ring'' we could only verify that the velocity map obtained by performing a similar analysis to that in \cite{CicBat18} appears compatible with the presence of the kinematic feature in this data-set, but it does not add significant information to refine its properties.

\begin{figure}[t]
\centering
\includegraphics[width=0.9\linewidth]{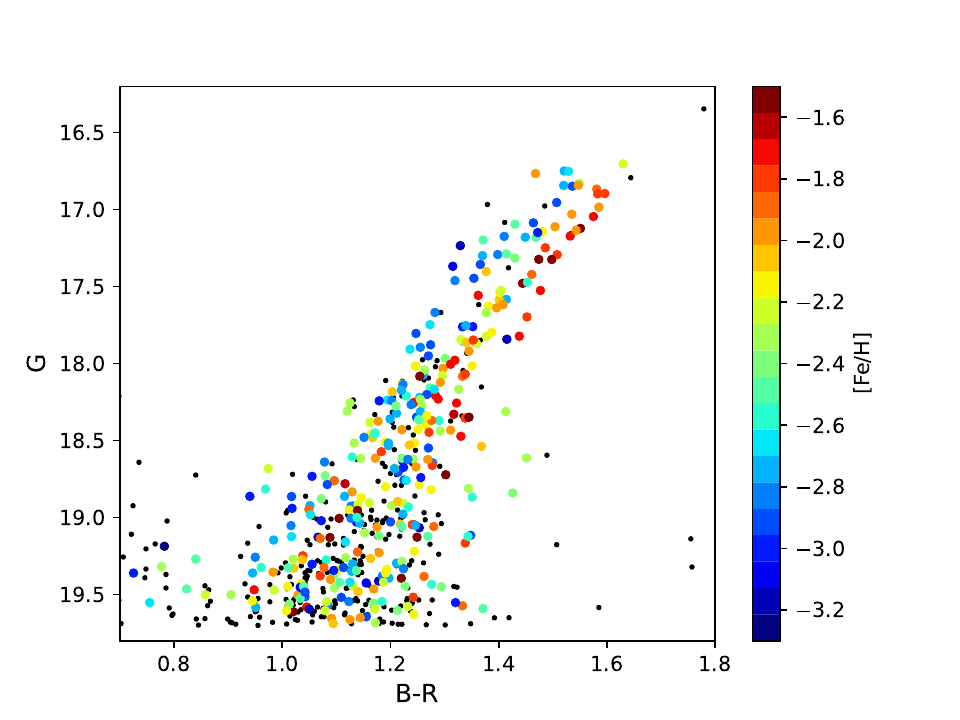}
\caption{The \gdr{3} reddening corrected Colour-Magnitude Diagram for the FLAMES Sextans dSph sample, with circles  
for those stars with  reliable FLAMES Ca~II triplet [Fe/H] measurements, colour coded by [Fe/H] as shown by the colour bar. The \gdr{3} sample without FLAMES spectra are shown as small black points.
 }
\label{cmd}
\end{figure}

\subsection{Chemo-kinematic properties}

It has already been established that the stellar component of several Milky Way classical dwarf spheroidal galaxies, including the Sextans dSph, can be described as the super-position of at least two stellar populations with different spatial distribution, mean metallicity and kinematics \citep[e.g.][]{Tolstoy04, Battaglia06, Battaglia08a, WalkPenn11, Pace2020}.

The presence of multiple chemo-kinematic components in the Sextans dSph was first explored in \citet{Battaglia11}. Guided by the presence of a clear spatial variation of the metallicity properties and by the fact that the MDF could be well fit by the sum of two Gaussians crossing at [Fe/H]$\sim -2.3$, the authors separated the sample into stars more MR than [Fe/H]$=-2.2$ and more MP than [Fe/H]$=-2.4$ and found that the \vlos\  dispersion profile of the MR population tended to be around 6 km s$^{-1}$ and slightly declining as a function of radius, while for the MP population it was approximately flat around $\sim$9.5 kms$^{-1}$ (apart from an inner cold point, with a velocity dispersion $\sim$1.4$\pm$1.2 kms$^{-1}$, discussed later).

Here we revisit the presence of multiple components using the same methodology that \cite{JM24} have recently applied to the Sculptor dSph, broadly based on \cite{WP11, Taibi18, Pace2020}. Here we simply investigate whether the combined information about the position, [Fe/H] and \vlos\ of member stars is best described by one population displaying a (linear) metallicity gradient or by the sum of two populations. In all cases, the surface density profile of the overall or individual stellar components is assumed to follow a Plummer profile, while the l.o.s. velocity distribution is assumed to be Gaussian. This analysis is applied to the combined sample of 303 stars selected to have S/N $>16$ { per pixel}, G < 19.7 and a \gaia+spec probability of membership $>$0.07. We correct the l.o.s. velocities for the relative motion between the Sextans dSph and the Sun.

The Bayesian parameter estimation for each model was obtained with pymultinest. We find that the 2-population model is strongly favoured, by 11 on the Jeffrey scale, over the single population model. 

The main properties of the two stellar components are summarized in Table~\ref{tab:Chemodynfit}. Similar to  previous findings and the general characteristics of multiple chemo-kinematic components in other Milky Way dSphs, the MR population is found to have a more concentrated spatial distribution and a colder velocity dispersion than the MP component. See Fig.~\ref{fig:MR_MP_components}. In a separate contribution, we will be analysing in detail the velocity dispersion profiles of the MR and MP population and the shape of the overall l.o.s. velocity distribution. At present, we do not see evidence of the inner cold-kinematic point found by \cite{Battaglia11} in the l.o.s. velocity dispersion profile in this new data set.

\subsection{\gdr{3} Colour-Magnitude Diagram}

In Fig.\ref{cmd} we plot the de-reddenend \gdr{3} photometry for the RGB stars that are \gdr{3} (P$_{mem} >0.07$) members, assuming E(B-V) $\sim 0.04$ \citep{Cicuendez18}), and where available FLAMES [Fe/H] measurements, colour coded from MR (red) to MP (blue) as shown on the colour bar in the plot. As in Paper~I for Sculptor dSph, we can see that the more MR stars are nicely aligned on the red side of the RGB and the more MP stars on the blue. Towards the tip of the RGB (G$<18$) there appear to be two distinct sequences. This is consistent with what has been found in Paper~I, and suggests that when identifications can be made precisely enough to identify two populations in the [Fe/H] histogram and kinematically in the velocity dispersion measures then they can also be identified in their colours on the RGB of the CMD. For fainter magnitudes (G$> 18.5$) it is not necessarily the case that the populations are more mixed, but that the errors in all the different measurements (\gdr{3} photometry and [Fe/H]) are increasing. 

\subsection{Palomar 3}
The Palomar~3 globular cluster was discovered long before the Sextans dSph, by A.G. Wilson in 1952, using photographic plates taken with the 48-inch Schmidt telescope at Palomar Observatory \citep{Abell55}. The first detailed study of the resolved stellar population was made by \cite{Burb58}, based on photographic plates from the 200-inch Hale telescope. They determined a distance of 100~kpc, very much in line with the modern determination of 95~kpc  \citep{BaumVasi21}. 
Palomar~3 has thus long been known to be one of the most distant globular clusters associated with the Milky Way. It is also one of a small number of Galactic globular clusters with "high-energy" dynamical properties \citep[e.g.][]{Massari19}. So its origin and association to any known structure in the Milky Way is unknown. According to \cite{VasiBaum21} Palomar~3
could be associated with Sagittarius
debris but the models are unreliable at these distances due to the lack of observational data.
It is clearly a MP, [Fe/H]$=-1.6$~dex \citep{Orto89, Armandroff92, Koch09} and relatively old system, although probably $1.5-2$~Gyr younger than the classical globular clusters \citep{Stetson99}. See Table~\ref{tablepal3} for an overview.

It appears noteworthy that the Sextans dSph and the Palomar~3 globular cluster, both of unknown origin, lie so close on the sky, and also at such similar distances, with Palomar~3 being $\sim$10~kpc more distant. Their \vlos\ are quite different, with Palomar~3 having a \vlos\ $= +92$km/s \citep{Peterson85, Koch09}, and so also their motion in the sky differs markedly (see Fig.~\ref{sextmems}). Likely members of Palomar~3 have the same proper motion and parallax signatures to Sextans (very close to zero), and so they were picked up naturally in our original target selection. We observed 6 stars likely associated to Palomar~3 and also a selection in the region around, see Table~\ref{pal3mems} for details of the 6 RGB stars we observed. 

The differences of the proper motion directions of the  Palomar~3 globular cluster and the Sextans dSph on the sky (Fig.~\ref{sextmems}) suggest that these two systems have no relation to each other (as has long been assumed given their very different \vlos\ ). The differences in 3D motions also makes it difficult for the Palomar~3 globular cluster to be bound to the Sextans dSph, unless Sextans is much more massive than it currently appears. 

\begin{figure}[ht]
\centering
\includegraphics[width=0.9\linewidth]{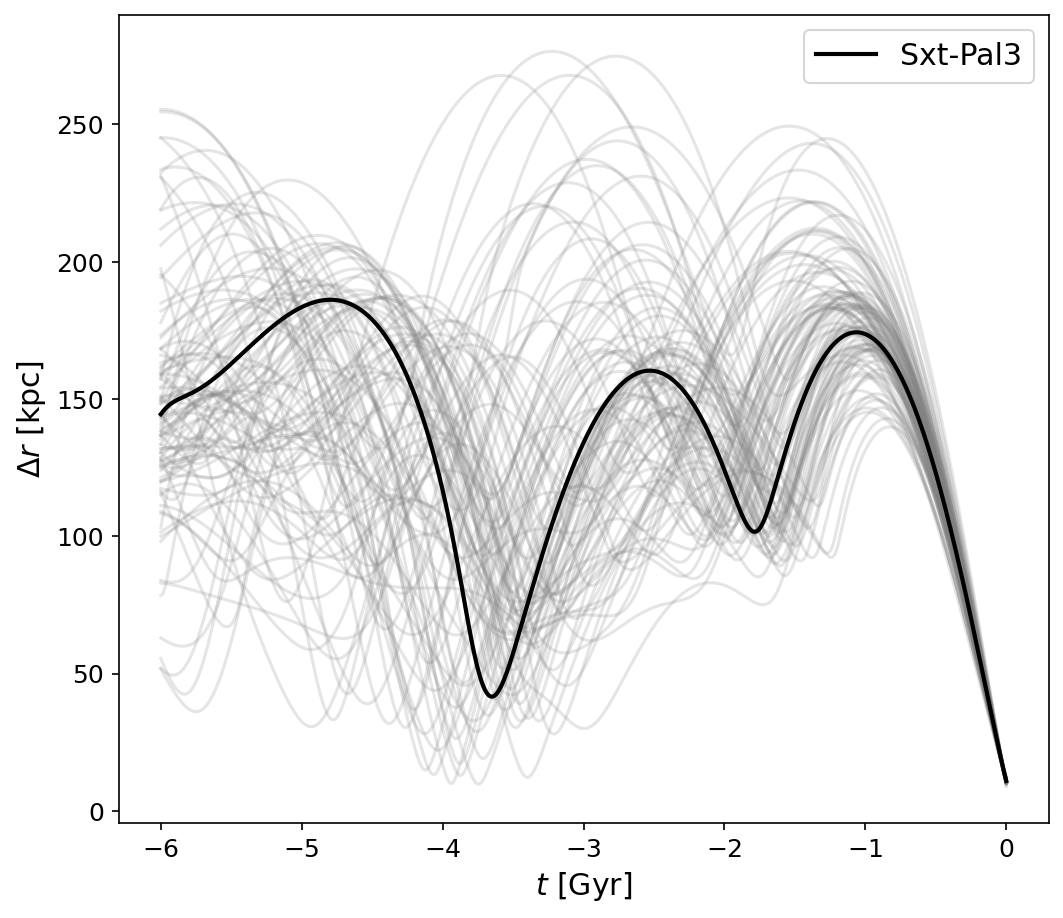}
\caption{From orbit integration for both Sextans dSph and Pal~3, using the \cite{Vasiliev21} potential, here is distance between Sextans and Pal3 going back in time.
 }
\label{orbit}
\end{figure}

We carried out a straigh forward  
orbit integration of the two systems 
based on their systemic proper motions, for a potential that includes the LMC potential as described in \cite{Vasiliev21}. In Fig.~\ref{orbit} we show how the distance between the Sextans dSph and
Pal~3 changes with time. In Appendix~C (Fig.~\ref{orbitsC1}) we show the details of the orbit integrations that went into this plot. It is clear that the uncertainties grow going further back in time, but given the assumptions we have used there is no evidence that Pal~3 and Sextans have any connection or signs of past interactions. It appears that they are currently closer than they have ever been.
The inferred apocentre and pericentre variations over the last 6~Gyr for Pal~3 and Sextans do overlap in Galactocentric distance but they follow different paths, with Sextans typically inhabiting more distant regions of the Milky Way halo with with a slower orbital trajectory. 

If these systems have been disrupted in some way in the recent past, which could perhaps explain the extremely low surface brightness of Sextans, then of course it is more difficult to be definitive about their orbital histories and a possible relation between these two systems. However, there is no obvious cause for any disruption, so any assumption that Sextans and/or Palomar~3 are out of equilibrium would be highly speculative. 
We also looked for any sign of Pal~3 stars more distant from the cluster in our FLAMES sample, based on the expected \vlos\ and proper motions, but did not find any.

The proper motion errors on Pal~3 are quite large, and the true mass of Sextans, let alone the Milky Way are quite uncertain, and also how they may change in the past, so there is potential for more detailed modelling, but this is well beyond the scope of this work.

\begin{table}
\caption{Mean properties of Palomar~3} 
\label{tablepal3}   

\centering     
\renewcommand{\arraystretch}{1.3}

\begin{tabular}{l c c}          
\hline\hline                     
Property & ~~value & ~~reference\\ 
\hline    
$\alpha$ & $151.383$& \\
$\delta$ &  $+0.072$ & \\
distance & $94.8 \pm 3$~kpc & [1] \\
\vlos & $95$ km/s & [2] \\
$[{\rm Fe/H]}$ & $-1.6$~dex & [2] \\
R$_0$ & 0.71 & [3] \\
$\mu_\alpha $ & $+0.086 \pm 0.060$ & [3] \\
$\mu_\delta $ & $-0.148 \pm 0.071$ & [3] \\
$\varpi$ & $-0.001 \pm 0.050$ & [3] \\
\hline                           
\end{tabular}


    \begin{tablenotes}
      \footnotesize
      \item $^{[1]}$ \citet{BaumVasi21}
      \item $^{[2]}$ \citet{Koch09}
      \item $^{[3]}$ \cite{VasiBaum21}
    \end{tablenotes}
\end{table}

\begin{table*}[t]
\centering     

\caption{Individual Stars observed in Palomar~3} 
\label{pal3mems}  
\renewcommand{\arraystretch}{1.3}
\begin{tabular}{l l c c c l}       
\hline                   
 Name & \gaia\ ID & Gmag & \vlos\ & [Fe/H]  &Notes\\ 
\hline    
DM354 & 3833327202956178304 & 18.8 & 96.14  & -1.57  $\pm$ 0.08 &  \\
DM353 & 3833327202955913472 & 18.4 & 92.85 & -1.56 $\pm$  0.06 &   K2 -1.56 $\pm$ 0.14\\
DM355 & 3833327379049493248 & 18.3 & 92.36 & -1.60  $\pm$ 0.05 &   K6 -1.61 $\pm$ 0.17 \\
DM685 & 3833327099876700288 & 18.0 & 92.80 & -1.43 $\pm$  0.05 &   K3 -1.58 $\pm$ 0.14 \\
NS46176 & 3833326996797478016 & 19.0 & 96.56 & -1.63 $\pm$  0.08  &  \\
DM351 & 3833326996797476864 & 17.7 & 96.31 & -1.48 $\pm$  0.03 &   \\

\hline                           
\end{tabular}
\tablefoot{K2, K3, K6, are stars observed with Magellan/MIKE by \cite{Koch09}, and the [Fe/H] they determined.}
\end{table*}

\section{Conclusions}

From a combination of new and old re-processed archival VLT/FLAMES LR8 data, we have obtained a uniform sample of [Fe/H], \vlos\ and \gdr{3} astrometric parameters for 312 individual member RGB stars, extending to (and beyond) the nominal tidal radius. 
This has allowed us to look again at the chemo-dynamical properties of the Sextans dSph. The main difference with previous results is that given the new more uniformly distributed, and more precisely defined ‘cleaner’ sample; thus the scatter is much better quantified and the features suggested seen in smaller data sets are typically less compelling, but still worth more detailed investigation and  modelling. 
We confirm that the Sextans dSph has complex \vlos\ and [Fe/H] distributions. The MR component is clearly concentrated in the centre of the galaxy. There are 
kinematic and [Fe/H] sub-structures in Sextans but  the sparsity of the tracers makes it difficult to know how robust they are without more detailed modelling. The wealth of information coming from the combination of spectroscopic metallicities from FLAMES and the exquisite \gdr{3} photometry of the sample of RGB stars in the Sextans dSph suggests a complex history. Larger samples of more precise high spectral resolution abundances would be beneficial to clarify the chemical differences in a more detailed and quantitative way.

\section*{Acknowledgements}

This work has made use of data from the European Space Agency (ESA) mission \gaia\ (\url{https://www.cosmos.esa.int/gaia}), processed by the \gaia\ Data Processing and Analysis Consortium (DPAC,
\url{https://www.cosmos.esa.int/web/gaia/dpac/consortium}). Funding for the DPAC has been provided by national institutions, in particular the institutions participating in the \gaia\ Multilateral Agreement.


\bibliography{et}

\clearpage
\onecolumn
\renewcommand{\thefootnote}{\fnsymbol{footnote}}

\begin{appendix}
\renewcommand{\thesection}{A}
\section{The Observations: individual VLT/FLAMES pointings} 

\setcounter{table}{0}
\begin{longtable}{l c l c c c c c c c c}
\caption{\label{tab:obs} LR8 observations of Red Giant Branch stars in the Sextans dSph with VLT/FLAMES/GIRAFFE. For each individual FLAMES pointing we provide the date of the observation, the ESO programme ID and the name we gave to each field as well as the RA and Dec of centre of the field, and the airmass (airm) and seeing monitor (DIMM) values. The exposure time (exp) is given. Also provided are the number of spectra in each pointing (spec) and the number of these that have a \vlos\ expected for membership in Sextans (vmem) and also the shift in the velocity of the entire field (vshift) with respect to our zero point, this shift ensure that all our observations are on the same velocity scale and that small effects between different observations are taken into account. }\\
\endfirsthead

\hline
No. & Date & Name & Prog ID  & Centre of field (J2000) & airm & DIMM &exp & spec & vmem & vshift \\
\hline
\hline
1&2003-12-20  & Cent-A  & 0171.B-0588 & 10 12 58.1  ~~  -01 38 05.5  & 1.34 & 0.62 & 2700 & 54 &40& 2.54\\
2&2003-12-20  & Corn15 & 0171.B-0588 & 10 11 30.4  ~~  -02 09 45.0  & 1.17 & 0.79 & 3600 & 54 & 4& 2.41\\
3&2003-12-22  & Corn1  & 0171.B-0588 & 10 14 47.4  ~~  -01 11 27.5  & 1.18  & 0.60  & 3800 & 55 &16& 2.03\\
4&2003-12-22  & Corn3  & 0171.B-0588 & 10 11 13.3  ~~  -01 12 01.9  & 1.43 & 0.93 & 4000 & 50 &5& 1.76\\
5&2004-01-01  & Corn13 & 0171.B-0588 & 10 14 52.3  ~~  -02 24 24.5  & 1.12 & 0.74 & 3600 & 47 &2& 1.49\\
6&2004-03-14  & f102 & 0171.B-0588 & 10 13 22.1  ~~  -01 21 53.2  & 1.29 & 0.58 & 3600 & 92 &16& 1.86\\
7&2004-03-15  & f104  & 0171.B-0588 & 10 11 21.6  ~~  -01 31 38.0  & 1.24 & 0.50  & 3600 & 68 &11& 2.85\\
8&2004-03-16  & f114  & 0171.B-0588 & 10 12 24.1  ~~  -01 58 43.5  & 1.56 & 0.80  & 4600 & 82 &21& 1.87\\
9&2004-03-17  & ext08	& 0171.B-0588 & 10 08 27.4 ~~   -00 52 01.6  & 1.37 & 0.74 & 3600 & 58 &0& 1.56\\
10&2004-03-17  & ext02	& 0171.B-0588 & 10 17 03.1   ~~ -01 01 49.8  & 1.43 & 0.55 & 3600 & 76 &3& 1.72\\
11&2004-03-18  & ext03	& 0171.B-0588 & 10 18 36.3  ~~  -00 24 21.9  & 1.33 & 0.89 & 3600 & 69 &1& 1.55\\
12&2004-03-19  & ext04	& 0171.B-0588 & 10 20 43.2  ~~  -00 06 51.4  & 1.29 & 0.93 & 3600 & 78 &0& 1.78\\
13&2004-03-20  & ext06	& 0171.B-0588 & 10 13 22.9  ~~  -00 26 51.1  & 1.52  & 0.77 & 4200 & 68 &1& 1.91\\
14&2004-03-20  & ext14	& 0171.B-0588 & 10 07 42.2  ~~  -03 13 27.8  & 1.31 & 0.95 & 2511 & 65 &0& 1.77\\
-&2004-03-21  & ext14	& 0171.B-0588 & 10 07 42.3  ~~  -03 13 27.8  & 1.42 & 0.50  & 3600 & 65 &0& 2.25\\
15&2004-03-21  & ext16	& 0171.B-0588 & 10 07 29.2  ~~  -01 40 19.7  & 1.23 & 0.80  & 4500 & 57 &1& 1.62\\
\hline
16&2008-05-25  & f112& 0060.A-9800 & 10 13 38.6  ~~  -02 00 35.7  & 1.57 & 0.82 & 3600 & 81 &23& 1.79\\
\hline
17&2019-03-11  & c1	& 0102.B-0786 & 10 13 44.5   ~~ -01 37 44.7  & 1.88 & 0.75 & 3600 & 73 &54& 2.52\\
18&2019-03-11  & c2	& 0102.B-0786 & 10 14 56.8  ~~  -01 37 11.3  & 1.38 & 0.76 & 3600 & 30 &23& 2.05\\
19&2019-03-11  & c3	& 0102.B-0786 & 10 16 13.0  ~~  -01 38 34.5  & 1.17 & 0.59 & 3600 & 11 &6& 2.34\\
20&2019-03-11  & c4	& 0102.B-0786 & 10 11 33.0  ~~  -01 50 09.1  & 1.09 & 0.53 & 3600 & 31 &28& 2.20\\
21&2019-03-11  & c5	& 0102.B-0786 & 10 12 48.0  ~~  -02 21 12.4  & 1.25 & 0.44 & 3600 & 12 &7& 1.98\\
22&2019-03-11  & c6	& 0102.B-0786 & 10 12 09.1  ~~  -01 17 05.7  & 1.12 & 0.49 & 3600 & 28 &25& 2.05\\
23&2019-03-12  & c7	& 0102.B-0786 & 10 13 42.9  ~~  -01 13 40.4  & 1.12 & 0.75 & 3600 & 33 &24& 2.33\\
24&2019-03-12  & c8	& 0102.B-0786 & 10 14 24.5  ~~  -00 42 25.4  & 1.18 & 0.54 & 3600 & 9  &5& 1.88\\
25&2019-03-12  & Pal3  & 0102.B-0786 & 10 05 29.0   ~  +00 00 33.6  & 1.11 & 0.38 & 3600 & 97 &0$^\ast$ & 2.31\\
26&2019-03-12  & c11	& 0102.B-0786 & 10 07 31.7  ~~  -00 03 01.1  & 1.45 & 0.67 & 3600 & 112&1& 1.75\\
27&2019-03-13  & c9	& 0102.B-0786 & 10 11 00.2  ~~  -02 09 30.2  & 1.22 & 1.43 & 3600 & 22 &11& 2.23\\
28&2019-03-13  & c12	& 0102.B-0786 & 10 08 45.0  ~~  -00 26 55.6  & 1.11  & 1.02 & 3600 & 113&0& 1.68\\
29&2019-03-13  & c13	& 0102.B-0786 & 10 05 27.9  ~~  -01 23 51.8  & 1.10 & 0.52 & 3600 & 107&0& 2.01\\
30&2019-03-13  & c14	& 0102.B-0786 & 10 10 05.8  ~~  -01 35 00.5  & 1.20 & 0.39 & 3600 & 110&6& 1.62\\
31&2019-03-13  & c15	& 0102.B-0786 & 10 07 34.0  ~~  -02 00 09.4  & 1.48 & 0.41 & 3600 & 111&0& 1.69\\
32&2019-03-14  & c16	& 0102.B-0786 & 10 03 43.9  ~~  -02 52 02.5  & 1.26 & 0.46 & 3600 & 7  &0& 1.88\\
33&2019-03-14  & c17	& 0102.B-0786 & 10 18 45.0  ~~  -01 11 29.8  & 1.15 & 0.61 & 3600 & 8  &2& 2.33\\
34&2019-03-14  & c19	& 0102.B-0786 & 10 16 47.7   ~~ -00 35 26.4  & 1.09 & 0.98 & 3600 & 10 &2& 2.00\\
35&2019-03-14  & c21	& 0102.B-0786 & 10 13 58.1  ~~  -03 18 39.6  & 1.30 & 0.91 & 3600 & 9  &1& 1.47\\
36&2019-03-14  & c24	& 0102.B-0786 & 10 15 07.2  ~~  -01 52 26.0  & 1.13 & 0.55 & 3600 & 11 &5& 1.84\\
37&2019-03-15  & c18	& 0102.B-0786 & 10 10 07.0  ~~  -02 58 20.5  & 1.08 & 0.49 & 3600 & 6  &2& 1.70\\
38&2019-03-15  & c20	& 0102.B-0786 & 10 18 00.0  ~~  -02 15 00.1  & 1.28 & 0.68 & 3600 & 8  &2& 1.84\\
39&2019-03-15  & c22	& 0102.B-0786 & 10 13 06.7   ~~ -00 02 55.5  & 1.17 & 0.54 & 3600 & 9  &1& 1.46\\
40&2019-03-15  & c23	& 0102.B-0786 & 10 22 36.2  ~~  -00 41 35.7  & 1.33 & 0.54 & 3600 & 7  &0& 1.51\\
41&2019-03-15  & c25	& 0102.B-0786 & 10 15 27.3  ~~  -01 13 54.1  & 1.13 & 0.73 & 3600 & 15 &8& 1.65\\
\hline
\label{table1}
\end{longtable}
\vspace{5pt} 
\noindent $^{\ast}$ There are 6 members of Pal~3.

\end{appendix}
\newpage

\begin{appendix}
\renewcommand{\thesection}{B}
\section{Comparing FLAMES Measurements with \cite{Walker23}} 
\label{app:walk}
\setcounter{figure}{0}
\renewcommand{\thefigure}{B.\arabic{figure}} 

\begin{figure}[ht]
\centering
\includegraphics[width=0.8\linewidth]{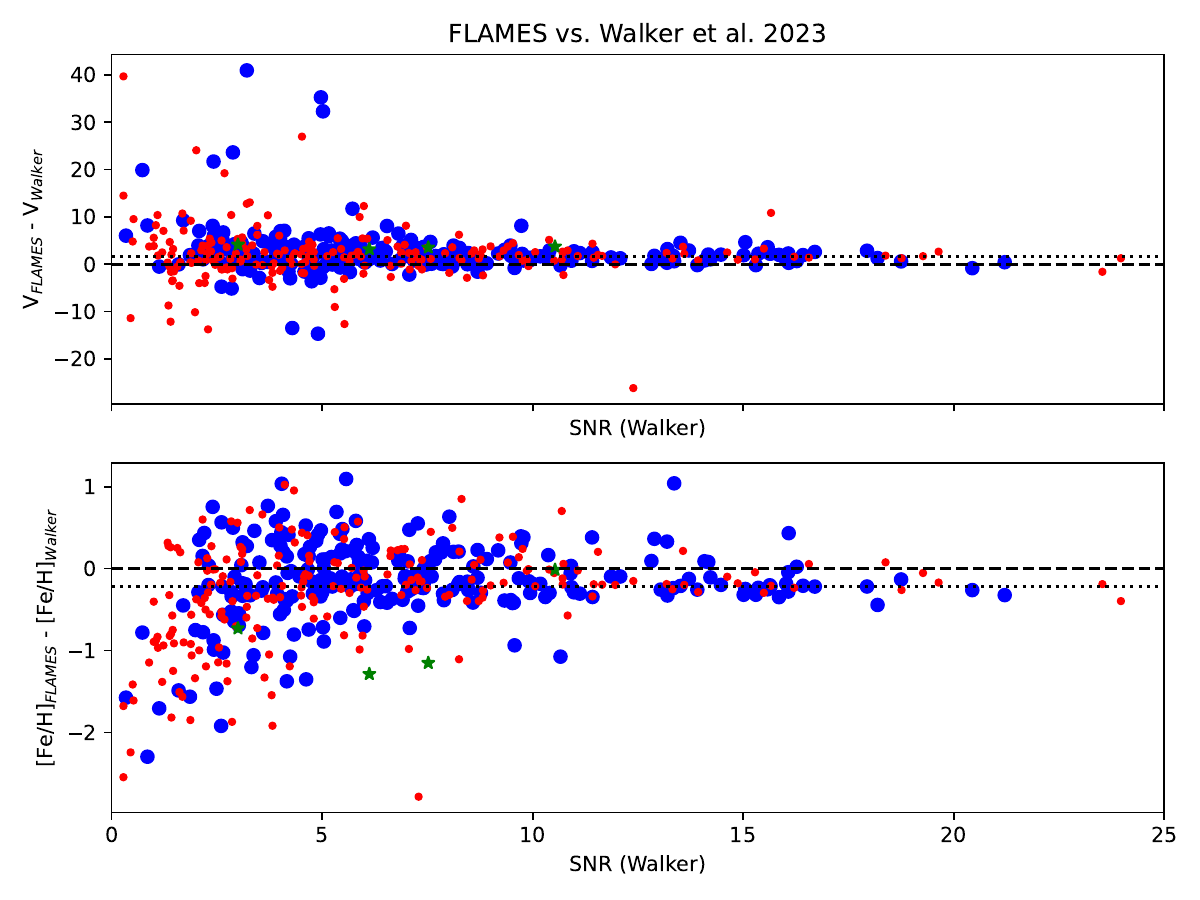}
\caption{ This is the comparison between \cite{Walker23} MMT/Hectochelle and  Magellan/M2FS spectroscopic results and our VLT/FLAMES results for the stars in common to both surveys. We compare with our final sample of members (G$<19.7$mag, S/N$>10$ { per pixel}).
In the upper panel, we plot the \vlos\ comparison of the two surveys and in the lower panel, [Fe/H]. The blue solid points are from the MMT and the red points are from the main M2FS survey. The green stars come from the medium resolution M2FS sample. 
The dashed lines are at zero and the dotted lines are the mean offset between the two surveys, (v$_{Walker} + 1.7$ km/s and [Fe/H]$_{Walker} +0.22$ dex.}
\label{velcomp}
\end{figure}

Our FLAMES sky coverage has a hole in the central region, and thus we explore combining our observations with the study of \cite{Walker23}, where many of these measurements were already presented in \cite{Walker09}. To make sure these observations over a different wavelength region made with different telescopes can be combined with our FLAMES observations we compare the \vlos\  and [Fe/H] of those stars that are observed in both surveys. This is shown in Fig.~\ref{velcomp}. We matched the Walker et al. catalogue of velocities and metallicities our final set of FLAMES members using the \gdr{3} ID. There are 442 matching measurements.

{ The S/N for the Walker measurements are not calculated in the same way as ours. There can be many reasons for this which we don't go into, but simply present the comparison between measurements made of the same stars and make our cuts based on this.} The differences between the Walker et al. \vlos\ measurements and ours for the same stars in Fig.~\ref{velcomp} look very stable, very comparable. 
There is a small offset with the mean velocities measured by Walker et al. of $+1.7$km/s, as shown by the dotted line in the upper panel of Fig.~\ref{velcomp}. This likely comes from using a slightly different zero point calibration. { The Walker et al. velocities were set to match the APOGEE DR17 results for the same stars, so this is an offset between our measurements and APOGEE.}
Some stars have quite large \vlos\ differences and these are potentially binary stars, where \vlos\ may differ significantly over the time between measurements, but the fact most of these offset measurements tend to have a low S/N suggests this is caused simply by measurement uncertainties.

The [Fe/H] measured by Walker et al. is determined using a different method to ours, namely { fitting a model  from a library of synthetic template spectra to the observed spectrum, providing both [Fe/H] and [Mg/Fe] determinations. Unfortunately, the resulting measurements [Fe/H] do not agree very well with ours. This is clearly seen in the lower panel of Fig.~\ref{velcomp}, where there is a considerable scatter, that is not uniform about a mean, especially at low S/N. The offset in the mean of the high S/N measurements is $+0.22$dex. We have not investigated this further, we have just opted not to include these measurements in our analysis.} 

\end{appendix}

\newpage
\begin{appendix}

\renewcommand{\thesection}{C}
\section{Orbit analysis of Sextans dSph and Pal~3} 
\label{app:orbit}
\setcounter{figure}{0}
\renewcommand{\thefigure}{C.\arabic{figure}} 

\begin{figure}[ht]
\centering
\includegraphics[width=0.8\linewidth]{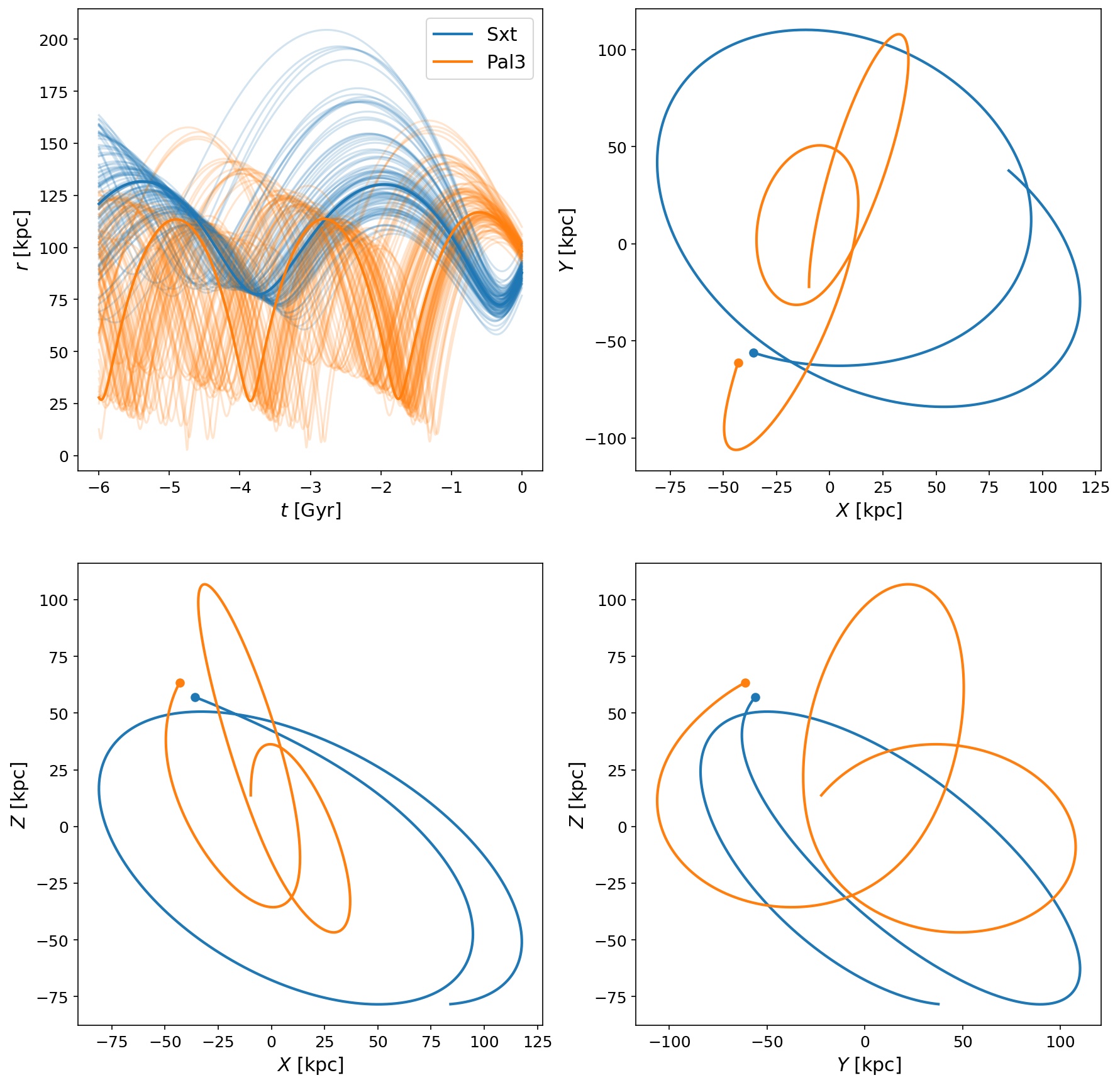}
\caption{ Details of the orbit analysis of Sextans and Pal~3. In the top left panel is the variation of the Galactocentric distance as a function of the look-back time for the two systems. In the other three panels is the orbital evolution in each of the planes of the xyz-space in Galactocentric coordinates.}
\label{orbitsC1}
\end{figure}

In Fig.~\ref{orbitsC1}, we show the orbital evolution of Sextans and Palomar~3 integrating backwards in time for 6~Gyr assuming the MW potential introduced in \citet{Vasiliev21}. This is a time-dependent potential that takes into account the passage of the LMC, with mass $1.5\times10^{11}M_\odot$ that entered the MW dark matter halo about 1.6~Gyr ago, and the acceleration induced by the reflex motion of the MW.
We used the code \texttt{Agama} \citep{Vasiliev2019} to perform the orbit integration. Orbital uncertainties were estimated from 100 Monte Carlo realisations by sampling the position-velocity vector of the inspected systems (shown for clarity only in the first panel of the figure).

\end{appendix}

\end{document}